\documentstyle[12pt,epsf]{article}

\def\yzero{\smash{\hbox{$y\kern-4pt\raise1pt\hbox{${}^\circ$}$}}}

\def\-{\hphantom{-}}
\def\ov{\overline}
\def\s2{\frac{1}{\sqrt2}}
\def\s22{\frac{1}{2\sqrt2}}

\def\beq{\begin{equation}}
\def\eeq{\end{equation}}
\def\beqa{\begin{eqnarray}}
\def\eeqa{\end{eqnarray}}
\newcommand{\bel}[1]{\begin{equation}\label{#1}}

\def\IF{\relax{\rm I\kern-.18em F}}
\def\II{\relax{\rm I\kern-.18em I}}
\def\IP{\relax{\rm I\kern-.18em P}}
\def\IR{\relax{\rm I\kern-.18em R}}
\def\inbar{\vrule height1.5ex width.4pt depth0pt}
\def\IC{\relax\hbox{\kern.25em$\inbar\kern-.3em{\rm C}$}}

\def\NN{{\cal N}}

\def\RR{{\cal R}}
   
\def\Dsl{\,\raise.15ex\hbox{/}\mkern-13.5mu D} 
\def\IZ{Z\kern-.5em  Z}
\def\cp#1{\relax\ifmmode {\IP\kern-2pt{}_{#1}}\else $\IP\kern-2pt{}_{#1}$\fi}
\def\e13{e^{2\pi i/3}}

\newcommand{\drawsquare}[2]{\hbox{%
\rule{#2pt}{#1pt}\hskip-#2pt
\rule{#1pt}{#2pt}\hskip-#1pt
\rule[#1pt]{#1pt}{#2pt}}\rule[#1pt]{#2pt}{#2pt}\hskip-#2pt
\rule{#2pt}{#1pt}}

\newcommand{\fund}{\raisebox{-.5pt}{\drawsquare{6.5}{0.4}}}
\newcommand{\antifund}{\overline{\fund}}

\begin{document}    
\pagestyle{empty}

\makeatletter
\@addtoreset{equation}{section}
\makeatother
\renewcommand{\theequation}{\thesection.\arabic{equation}}
\title{Brane Boxes and Branes on Singularities}

\author{Amihay Hanany, ~~~Angel M. Uranga\\ {\tt hanany, uranga@ias.edu}\\
\\School of Natural Sciences\\
Institute for Advanced Studies\\
 Princeton, NJ 08540, USA}
 
\maketitle

\vspace{-4.5in}\hspace{4in} IASSNS--HEP--98/45

\hspace*{4in} hep-th/9805139

\vspace{4.5in}

\begin{abstract}
Brane Box Models of intersecting NS and D5 branes are mapped to D3 
branes 
at $\IC^3/\Gamma$ singularities and vise versa, in a setup which
gives rise to $N=1$ supersymmetric gauge theories in four dimensions.
The Brane Box Models are constructed on a two-torus.
The map is interpreted as T-duality along the two directions of the torus.
Some Brane Box Models contain NS fivebranes winding around $(p,q)$ cycles 
in the torus, and our method provides the geometric T-dual to such objects. 
An amusing aspect of the mapping is that T-dual configurations are 
calculated
using $D=4$ $N=1$ field theory data.
The mapping to the singularity picture allows the geometrical 
interpretation of all the marginal 
couplings in finite field theories. This identification is further 
confirmed using the AdS/CFT correspondence for orbifold theories. The 
AdS massless fields coupling to the marginal operators in the boundary 
appear as stringy twisted sectors of $S^5/\Gamma$. 
The mapping for theories which  are non-finite requires the introduction 
of fractional D3 branes in the singularity picture. 
\end{abstract}

\setcounter{page}{1}
\pagestyle{plain}
\renewcommand{\thefootnote}{\arabic{footnote}}
\setcounter{footnote}{0}
\newpage
\tableofcontents
\newpage
\section{Introduction}

Brane Boxes are good objects for studying aspects of $N=1$ supersymmetric
chiral gauge theories in four dimensions and their corresponding dimensional
reductions.
These objects correspond to a simple generalization of the idea in \cite{hw}.
In \cite{hw}, a Dp brane is stretched in between a pair of two NS branes with
one direction being finite. The low energy effective field theory which lives
on the Dp brane is therefore a $p+1$ dimensional theory compactified on an
interval with length given by the distance between the two NS branes.
For small enough interval, the field theory on the brane is $p$ dimensional.
In addition, the NS branes impose boundary conditions which remove some of the
fields which live on the D-brane as well as reduce the supersymmetry by half.
The resulting theory is a $p$ dimensional theory with 8 supercharges and $p$
is less or equal to 6. When two brane intervals touch each other there are
additional massless multiplets. with 8 supercharges, these correspond to
bi-fundamental hypermultiplets which transform under the two gauge groups which
leave on the two brane intervals.

Brane boxes \cite{hz} are a generalization of this idea to a two dimensional
interval. A Dp brane is stretched in between two pairs of two NS branes with
two directions being finite. The low energy effective field theory which lives
on the Dp brane is therefore a $p+1$ dimensional theory compactified on two
intervals with each length given by the distance between the two corresponding
NS branes.
For small enough interval, the field theory on the brane is $p-1$ dimensional.
The NS branes impose boundary conditions which remove more fields
which live on the D-brane as well as reduce the supersymmetry by a further half.
The resulting theory is a $p-1$ dimensional theory with 4 supercharges and $p$
is less or equal to 5.
There are additional states which are massless when two brane boxes touch.
With 4 supercharges, these correspond to bi-fundamental chiral multiplets
which transform as fundamental and anti-fundamental of the corresponding
gauge groups. A superpotential term is present when three such boxes meet.
Three chiral multiplets form into a dimension 3 singlet operator which
contributes to the superpotential.

Using these rules, a large class of finite chiral $N=1$ supersymmetric gauge
theories were constructed in \cite{hsu}. The beta functions of the gauge
theories are directly related to the bending of the branes. Models with
vanishing beta function correspond to configurations with no bending.
An important class of models was introduced which are not finite but still
have some exact marginal operators. Such models obey a ``sum of diagonal 
rule''.
This rule was independently considered in \cite{gg} from a different point of
view and is discussed in the next paragraph.

The bending of the branes remains an open question. For weak coupling the
NS branes are much heavier than the D branes and are weakly affected by them.
For finite string coupling, NS branes which have no balance of D branes ending
on them start to bend. It is an important problem to understand such a
bending. This information will provide us an understanding of the dynamics
of strongly coupled chiral gauge theories.
A first step towards understanding such a bending was done in \cite{gg}.
A constraint on the possible ranks of four gauge groups which meet on a
vertex of four brane boxes was derived. This condition requires smoothness
of asymptotic bending. In some cases like supersymmetric Yang-Mills theories
we need not expect such smoothness and so there is a large class of gauge
theories for which this condition is too restrictive. Nevertheless, models
which do obey these conditions are nicely behaved models which can teach
us a lot on the dynamics of the corresponding gauge theories.
As a simple example, brane configurations which satisfy these conditions are
anomaly free \cite{gg}.
Other aspects of the brane boxes and their beta function were considered in
\cite{ab}.

In a different approach, initiated in \cite{dm}, branes on singularities
were analyzed and were shown, in some cases, to have gauge theories which are
identical to models derived in the approach discussed above.
In this paper we study this correspondence by showing that the two approaches
are related by T-duality along two directions.
A similar analysis for theories with 8 supercharges appears in 
\cite{lust}. In theories with four supercharges, the constructions of 
\cite{lpt1,lpt2} were shown to be equivalent to brane box construction in 
\cite{hz}, using T-duality along one direction.

The paper is organized as follows. In section two we present two different
constructions which lead to four dimensional $N=1$ supersymmetric gauge
theories. We start by reviewing the Brane Box Models of \cite{hz}. Then
we proceed to review the branes at singularities of
\cite{dm,dgm,ibanez,ks,lnv}.
Other related work appears in \cite{bkv,bj,k1,k2}.
We describe the rules which lead to the calculation of the gauge groups,
the matter content and the particular cubic interaction terms in the
superpotential.
One important set of Brane Box models has non-trivial identifications when
going around the circle on which they are defined.
These models are reviewed in this section to be discussed as the general models
in the next sections.
The models of branes at singularity which are studied in this paper correspond
only to Abelian discrete subgroups of $SU(3)$. Non-Abelian subgroups are
harder to map to Brane Box Models and are left for further study.

In section three, we present a constructive method of building a Brane Box
Model from a given singularity model.
The construction is formal and serves as a prepartion to a T-duality relation
between the two types of setups.

In section four, we use T-duality to construct a brane at singularity from a
given Brane Box Model.
This method allows us to calculate T-dual pairs between NS branes which wrap a
torus in various ways and a singularity of the form $\IC^3/\Gamma$, with
$\Gamma$ a discrete subgroup of $SU(3)$.
This is one of the amusing aspects of the present paper in which four
dimensional $N=1$ supersymmetric gauge theories are used to calculate
non-trivial dual pairs in Type II superstring theory.

In section five we discuss the counting of marginal couplings for the general
class of models introduced in section two. These parameters receive a 
geometric interpretation in the brane box picture as distances between NS 
fivebranes, and Wilson lines around compact directions. In the singularity 
picture they correspond to integrals of the Type IIB RR and NS two-forms over 
two-cycles implicit in the singularity. As a check of this identification, 
we use the AdS/CFT correspondence, and relate the four-dimensional $\NN=1$ 
finite theories to string theory on $AdS_5\times S^5/\Gamma$. The massless 
scalar fields propagating in $AdS_5$ which couple to the marginal 
operators in the boundary are seen to arise from stringy twisted sectors 
of $S^5/\Gamma$. This identification of the gauge theory parameters in the 
brane pictures allow us to make some qualitative statements about the 
strong coupling regime of the gauge theories.

In section six we discuss models which are not conformal field theories.
The different gauge theories in the Brane Box models are mapped to fractional 
branes living on singularities.
This leads us to a special class of models constructed from ``sewing'' three
different $N=2$ models into a general $N=1$ model.
Each $N=2$ model has a Coulomb branch which becomes part of the Higgs 
branch of
the $N=1$ model. Using the $N=2$ beta functions, the one loop beta function
for the $N=1$ model is calculated and is given an interpretation in the brane
picture.

\section{The constructions}

\subsection{Overview of the brane box configurations}

The $N=1$ models we will be considering are constructed in a
brane setup, in the spirit of \cite{hw}, which was described in detail
in \cite{hz}.  The description here will be short and further details
are contained in \cite{hz}.

We are working in Type IIB superstring theory with the following set
of branes.
\begin{itemize}
\item NS branes along 012345 directions
\item NS$'$ branes along 012367 directions
\item D5 branes along 012346 direction.
\end{itemize}
The D5 branes will be finite in two of the directions, 4 and 6; their
low-energy effective world volume theory is 3+1 dimensional.  The
presence of all branes breaks supersymmetry to 1/8 of the original
supersymmetry, and thus we are dealing with $N=1$ supersymmetry (4  
supercharges) in four dimensions.  The 4 and 6 directions are circles with
radii $R_4$ and $R_6$ respectively.

A generic configuration consists of a grid of $k$ NS branes and $k'$
NS$'$ branes in the 46 plane. This divides the 46 plane into a set of
$k k'$ boxes. In each box, we can place an arbitrary number of D5
branes. Let $n_{i,j}$ denote the number of D5 branes in the box $i,j$,
$i=1, \ldots, k$, $j=1, \ldots, k'$.  In the following, indices will
denote variables in a periodic fashion: an index $i$ is to be
understood modulo $k$ and an index $j$ is to be understood modulo
$k'$.  Thus a model's gauge and matter content is specified by the
numbers $k$ and $k'$ and the set of numbers $\{n_{i,j}\}$.

\begin{figure}
\centering
\epsfxsize=3.5in
\hspace*{0in}\vspace*{.2in}
\epsffile{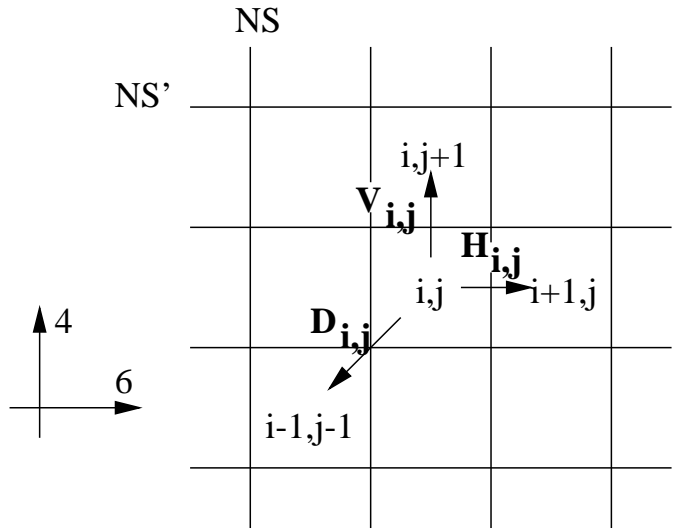}
\caption{\small Conventions for denoting the chiral multiplets which
are in the fundamental of the group $SU(n_{i,j})$ and
in the antifundamental of an adjacent group.}
\label{fig:multiplets}
\end{figure}

The gauge group is $\prod_{i,j}SU(n_{i,j})$.
The matter content of the model consists of three types of $N=1$ chiral
representations.  They will be denoted as $H_{i,j}$, $V_{i,j}$ and
$D_{i,j}$, corresponding to the horizontal, vertical and diagonal
multiplets which arise in the brane system (see the details in
\cite{hz}). $H_{i,j}$ transforms in the $(\fund,\antifund)$ of
$SU(n_{i,j})\times SU(n_{i+1,j})$, $V_{i,j}$ transforms in the
$(\fund,\antifund)$ of $SU(n_{i,j})\times SU(n_{i,j+1})$ and $D_{i,j}$
transforms in the $(\fund,\antifund)$ of $SU(n_{i,j})\times
SU(n_{i-1,j-1})$. Figure \ref{fig:multiplets} shows the conventions
for denoting the multiplets.

\begin{figure}
\centering
\epsfxsize=3in
\hspace*{0in}\vspace*{.2in}
\epsffile{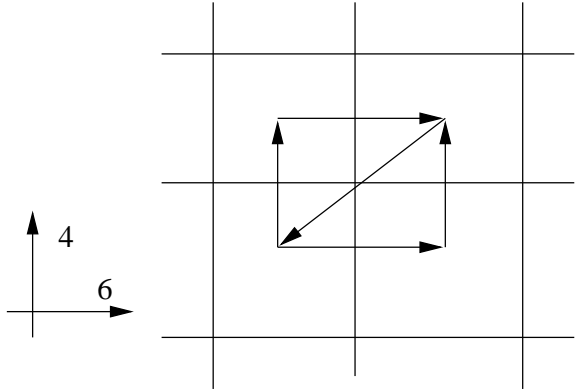}
\caption{\small The two superpotential couplings at each corner are 
represented by an oriented triangle of arrows.}
\label{fig:superpot}
\end{figure}

The superpotential in these models is calculated using the rules
described in \cite{hz}.  It is given by
\bel{HVD}
W=\sum_{i,j}H_{i,j}V_{i+1,j}D_{i+1,j+1}
-\sum_{i,j}H_{i,j+1}V_{i,j}D_{i+1,j+1}.
\eeq
The first term corresponds to lower triangles of arrows and the second
term corresponds to upper triangles of arrows in the notation of
\cite{hz}, as shown in figure \ref{fig:superpot}.  Note the relative
minus sign between the two terms.

\subsubsection{Models with non-trivial identifications}

The brane box models described above are defined on a torus in the 46 direction
in which the NS branes are trivially identified when going around any of 
the circles.
In this section we will review brane box configurations which have non-trivial
identifications once going around one of the circles of the torus.
The simplest example of this type of models was given in figure 7 of \cite{hsu}.

The construction goes as follows.
For any integers $k$ and $k'$ we can form a $k\times k'$ box model as in the
models with trivial identification.
Without loss of generality, we can assume that along one of the directions of
the torus the NS branes are identified trivially. Let us choose it to be the
4 direction.
Let $p$ be an integer between 0 and $k$. We place a $k\times k'$ box model
on top of another such box shifted to the left by $p$ boxes.
This gives NS$'$ branes which are trivially identified when going around the
4 circle. On the other hand the NS branes are identified non-trivially.
These models will be discussed further in section 4.2.
 
$p=0$ corresponds to the models with trivial identification which are discussed
in the previous subsection.

\subsection{Marginal Couplings}

The gauge couplings of the various gauge groups receive contributions from
various sources. The simplest contribution is expressed in terms of the
positions of the NS branes in
the $x^6$ direction and the position of the NS$'$ branes in the $x^4$
direction.  There are $k$ positions $x_6^i$ and $k'$ positions
$x_4^j$.  Correspondingly, the $x_6$ direction is divided into $k$
intervals with lengths $a_i=x_6^i-x_6^{i-1}$, such that
$\sum_ia_i=R_6$.  The $x_4$ direction is divided into $k'$ intervals 
of length $b_j=x_4^j-x_4^{j-1}$, such that $\sum_jb_j=R_4$.  The simplest
contribution to the gauge
coupling $g_{i,j}$ for the group $SU(n_{i,j})$ is given by
\bel{gauge}
{1\over g_{i,j}^2} = {a_i b_j \over g_s l_s^2}.
\eeq
The $kk'$ gauge couplings are not all independent. In equation 
(\ref{gauge})
they are given by
$k+k'-1$ parameters corresponding to the positions of the NS and NS$'$
branes.  Two positions can be set to zero by the choice of origin in  
the 46 directions, but the area of the 46 torus gives one more
parameter. The couplings do not depend on the ratio between the two
radii of the torus. As we will see later, using the mapping to the
branes on singularities, the field theories often
have more than $k+k'-1$ parameters, indicating that we have not
identified all of the contributions to these couplings.

The theta angles of the gauge theories receive various
contributions. Let $A_i$ be the gauge field on the world volume of the
$i^{th}$ NS brane and $A'_j$ be the gauge field on the world volume of
the $j^{th}$ NS$'$ brane. Since the dimensions 4 and 6 are compact,
there can be non-zero Wilson lines of $A_i$ along 4, and of $A'_j$ 
along 6.  Let $R_{i,j}$ denote the area in the 46 direction which is
bounded by the pair of NS branes and NS$'$ branes. The theta angle for
the $i,j$ group depends on the line integral of the different gauge
fields along the boundary of $R_{i,j}$. Schematically, 
\bel{theta}
\theta_{i,j}= \int_{R_{i,j}}B + \int_{a_i}(A'_{j-1}-A'_{j}) +
\int_{b_{j}}(A_{i}-A_{i-1}).
\eeq
where $B$ is the RR two form of Type IIB superstring theory. The
contributions from the gauge fields are required for the invariance of
$\theta_{i,j}$ under gauge transformations of $B$.  Were this the  
entire story we would again have $k+k'-1$ parameters.  However,
invariance under gauge transformations of the one-forms require that
additional terms be added to this expression involving axion-like
fields living at the intersections of the NS and NS$'$ branes.

In general, when quantum effects are taken into account these quatities 
run accordingly with the renormalization group. However, the brane 
configurations allow for a simple construction of $N=1$ theories which 
have some marginal couplings, {\em i.e.} a submanifold of renormalization 
group fixed points in the space of couplings. For example, the field 
theory analysis in \cite{hsu} showed that models which satisfy the 
``sum of diagonals rule,''
\beq
n_{i,j}+n_{i+1,j+1}=n_{i+1,j}+n_{i,j+1}
\label{sodr}
\eeq
give rise to non-finite models which have some marginal operators.
This condition was discussed in the context of brane bending in \cite{gg}.
We will discuss issues related to this condition in section 6.
The simplest models verifying these conditions are those in which all 
$n_{i,j}$ are equal. They give rise to exactly finite theories, in which 
no parameter is renormalized.

Let us now discuss the number of independent parameters which contribute 
to the gauge couplings. We claim that there are actually
$$
k+k'+r-2
$$
such couplings, where $r$ is the greatest common divisor of $k$ and $k'$,
$r=\gcd(k,k').$
For models with non-trivial identifications (of the type described in 
section 2.1.1) the number of marginal couplings generalizes to
$$
k + \gcd(k,p) +\gcd(k,k'+p)-2
$$
This counting follows from the field theory analysis performed in 
\cite{hsu}. From the point of view of the brane box construction,
we will give few arguments which support this claim. Further evidence will 
be provided in Section~5, using the T duality with the picture 
of branes at singularities.
 
First consider from a field theory point of view the asymmetry in the
construction in terms of branes. There are matter fields which come from
horizontal, vertical and diagonal arrows, however, the parameters which 
are seen
in equation (\ref{gauge}) correspond only to horizontal and vertical 
distances.
There are no parameters which correspond to diagonal fields.
To make the discussion more clear, let us center on a $k\times k'$ box 
model with trivial identifications, and let us define the following 
quantities.
\bel{ll}
h_i=\sum_{j=1}^{k'}{1\over g^2_{i,j}},\qquad
v_j=\sum_{i=1}^{k}{1\over g^2_{i,j}},\qquad
d_l=\sum_{diagonals}{1\over g^2_{i,j}}.
\eeq
The sum over diagonals means that one starts with some box and then 
proceeds
along the diagonal arrows until coming back to the first box. There are 
$r$
different such diagonals which correspond to $r$ different parameters, 
$d_l$
and each sum contains $kk'/r$ summands such that each gauge coupling 
appears
exactly once in one of the parameters $d_l$.
We claim that these quantities form a set of independent paramters which
give rise to the various gauge couplings.

The asymmetry becomes more clear when we consider a different brane box
construction which gives rise to the same field theory.
For a given field theory, as above, we can always rename what we call the 
fields
$H, V$ and $D$ by any permutation, say $D, V$ and $H$.
There are actually 6 such choices of renaming the fields.
While this is a formal process from a field theory point of view which 
does not
change the matter content, it is a crucial difference from the point of 
view
of brane box construction.
In the example of permutation chosen, fields which in the original 
construction
come from horizontal arrows, come in the permuted construction from 
diagonal
arrows and vise versa. Vertical fields on the other hand remain vertical.

\begin{figure}
\centering
\epsfxsize=6in
\hspace*{0in}\vspace*{.2in}
\epsffile{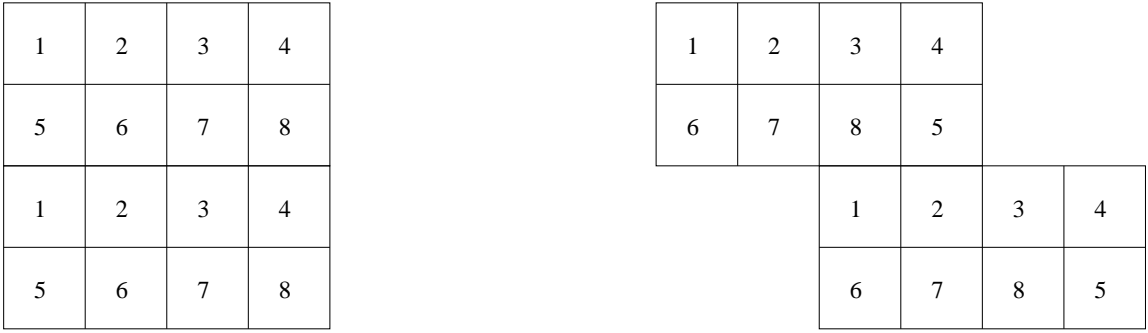}
\caption{\small A permutation on the types of fields. In this particular 
example the
$H, V, D$ fields are transformed into $H, D, V$ fields, respectively.
The $4\times 2$ box model with trivial identifications is mapped to the
$4\times 2$ model with a nontrivial identification with a horizontal
jump by $p=2$. The numbers in the boxes indicate labels of the boxes.}
\label{fig:fourtwo}
\end{figure}

The example in figure \ref{fig:fourtwo} shows how to construct such a brane
box permutation. A systematic way of constructing the permuted box model from
the original one is as follows. We pick a chain of periodic horizontal boxes.
We order the boxes in a new order which is specified by the permutation.
 From each box of the original horizontal chain we pick a chain of periodic
vertical boxes. We put these boxes in the new direction specified by the
permutation. The permutation on diagonal chain of boxes is already build into
the new setup. The new box model is done.

Let us count parameters in the original model and its corresponding
permuted model.
In figure \ref{fig:fourtwo} a, there are 4 vertical lines and there are 2
horizontal lines. (Here we mean in a unit box of the torus).
In addition there are 2 independent closed chains of
diagonals (given by {4,7,2,5} and {8,3,6,1}). The number of diagonals is
given for general $k$ and $k'$ by $r=\gcd(k,k')$.
In the permuted model of figure \ref{fig:fourtwo} b, there are 2 vertical lines
and 2 horizontal lines. On the other hand, there are 4 diagonal chains.
In both cases, we will count 4+2+2-2=6 marginal parameters for the field theory.
We see, as expected from the permutation, that the number of vertical and
the number of diagonal parameters are exchanged, while the number of 
horizontal lines is not changed.

In any of the models, the diagonal parameters are not easily visible but
the vertical parameters are visible. A symmetry of the construction from a 
field theory point of view, thus implies that indeed we have counted 
correctly the set of the marginal paramaters of the model. This counting 
will be useful for the identification of parameters in the singularity 
picture in Section~5.

\subsection{Overview of the branes at singularity}

\subsubsection{The spectrum}

The dynamics of four dimensional $\NN=4$ gauge theories can be 
studied by realizing them in the world-volume of parallel Type IIB 
D3-branes. Let us state, for concreteness, that such world-volume 
spans the coordinates 0123. In this context many properties of the gauge 
theory are usefully related to properties of string theory and the brane 
dynamics. For instance, the $SU(4)_R$ R-symmetry appears as 
the $SU(4)\approx SO(6)$ rotation group on the transversal coordinates 
456789. The gauge coupling constant is given by the string coupling, and 
Montonen-Olive duality is realized as the ten-dimensional $SL(2,\IZ)$ of 
Type IIB superstring theory. 

In \cite{lnv} this idea was extended by introducing a family 
of four-dimensional gauge field theories with reduced supersymmetry, which 
can be realized in the world-volume theory of D3 branes sitting at a 
singular point. This singularities are locally described as $\IC^3/\Gamma$. 
Here $\IC^3$ corresponds to the coordinates transverse to the D3 brane, 
namely 456789, and $\Gamma$ is a discrete subgroup of $SO(6)\approx SU(4)$. 
Since $\Gamma$ acts on the R-symmetry of the theory, the amount of 
surviving supersymmetry depends on this action. Theories with $\NN=2$ 
supersymmetry are obtained when $\Gamma \subset SU(2)$, $\NN=1$ 
(generically chiral) gauge theories appear if $\Gamma\subset SU(3)$, and 
non-supersymmetric theories correspond to $\Gamma$ being a generic 
subgroup of $SU(4)$.

The spectrum of the resulting theory can be analyzed using the techniques 
developed in \cite{dm}. We review the result for $\Gamma$ an {\em Abelian} 
subgroup of $SU(3)$, since we will be interested in this particular 
family of $\NN=1$ theories.

Let $|\Gamma |$ denote the order of $\Gamma$. A configuration of $N$ D3 
branes at the orbifold can be studied on the covering flat space by considering 
$N$ groups of $|\Gamma |$ D3 branes. $\Gamma$ acts on the set of $N |\Gamma |$ 
Chan-Paton factors as $N$ copies of the regular ($|\Gamma |$-dimensional) 
representation $\RR_{\Gamma}$, {\em i.e.} $\RR_{C.P.}=N \RR_{\Gamma}$
(Other embeddings of the Chan-Paton factors and their interpretation will 
be discussed in Section~6). When $\Gamma$ is Abelian it has $|\Gamma |$ 
unitary irreducible representations $\RR_I$, all of which 
are one-dimensional. The regular representation is reducible and 
decomposes as $\RR_{\Gamma}=\bigoplus_I  \RR_I$. One must also define 
how $\Gamma$ acts on $\IC^3$ to form the quotient singularity; this is 
specified by a (faithful) three-dimensional representation, which has a 
decomposition in irreducible representations as:
\beq
{{\bf 3}}\; =\; \RR_{A_1} \oplus \RR_{A_2} \oplus \RR_{A_3}
\eeq

In order for ${\bf 3}$ to be a representation of $SU(3)$ rather than of 
$U(3)$ there is a requirement on the choice of the representations 
$\RR_{A_i}$, whose statement 
is easier by noticing the following fact.
The set of irreducible representations forms an Abelian group (isomorphic 
to $\Gamma$) with respect to the tensor product of representations. We 
write the group law as
\beq
\RR_I \otimes \RR_J \; \equiv \;  \RR_{I\oplus J}
\eeq
Using this additive notation on the indices of the irreducible 
representations, the trivial 
representation is denoted $\RR_0$, and $\RR_{-I}$ denotes the inverse 
element of $\RR_{I}$. The requirement on the representation ${\bf 3}$ can 
be stated as $\RR_{A_1}\otimes \RR_{A_2}\otimes \RR_{A_3}=\RR_0$, or 
equivalently as $\RR_{A_3}=\RR_{-A_1-A_2}$.

The construction in \cite{lnv} results in the following spectrum. The 
gauge group \footnote{The $U(1)$ gauge fields (but one) are not expected 
to appear in the low energy dynamics of the D-branes. A possibility 
is 
that they are broken by a Green-Schwarz mechanism \cite{afiv}, as happens
in certain six dimensional models \cite{dm,ber,intri}.}
 contains a factor $SU(N)$ per irreducible representation of 
$\Gamma$, so the group 
is $\prod_{I} SU(N)_I = SU(N)^{|\Gamma |}$.
The chiral matter is found by computing the products
\beq
{\bf 3} \otimes \RR_I = \RR_{I\oplus A_1} \oplus \RR_{I\oplus A_2} \oplus 
\RR_{I-A_1-A_2} 
\eeq
There are three kinds of chiral multiplets, which are associated to the 
three complex planes transverse to the D3 branes. We will denote them by 
$\Phi_{I,I\oplus A_i}$, for $i=1,2,3$. The fields $\Phi_{I,I\oplus A_1}$ 
transform in the $(\fund,\antifund)$ of $SU(N)_I\times SU(N)_{I\oplus 
A_1}$, $\Phi_{I,I\oplus A_2}$ transform in the $(\fund,\antifund)$ of 
$SU(N)_I\times SU(N)_{I\oplus A_2}$, and $\Phi_{I,I\oplus A_3}$ 
transforms in the $(\fund,\antifund)$ of $SU(N)_I\times SU(N)_{I\oplus 
A_3}$.
Notice there are three such fields per irreducible representation of $\Gamma$.

Finally, for each $I$ there are two cubic terms in the superpotential, 
which takes the form
\beqa
W \; = \; \sum_{I} \big[ \Phi_{I,I\oplus A_1} \,  
\Phi_{I\oplus A_1,I\oplus A_1\oplus A_2}\,  
\Phi_{I\oplus A_1\oplus A_2,I} \; - \; \Phi_{I,I\oplus A_2} \,  
\Phi_{I\oplus A_2,I\oplus A_2\oplus A_1}\,  \Phi_{I\oplus 
A_2\oplus A_1,I}\big]
\label{superp}
\eeqa

Before going further in the discussion of these models and their relation 
to the brane box configurations, it will be useful to discuss a few 
examples.

\medskip

\subsubsection{Examples}

In the following we describe the spectrum for D3 branes on 
some singularities. Since the case of $\NN=2$ theories (corresponding to 
an $A_k$ ALE singularity), has been largely discussed 
in the literature \cite{dm}, we will mention only $\NN=1$ models.

\medskip

{\bf i)  $\IC^3/Z_3$}

Consider a $\IC^3/\IZ_3$ singularity, where the generator $\theta$ of 
$\IZ_3$ acts on $\IC^3$ as 
\beq
(z_1,z_2,z_3) \to (\e13 z_1,\e13 z_2,\e13 z_3).
\label{actionz3}
\eeq
This is the only choice of the representation ${\bf 3}$ that leaves 
unbroken $\NN=1$ supersymmetry (and not $\NN=2$).
The group $\Gamma=\IZ_3$ has three one-dimensional irreducible 
representations $\RR_{I}$, $I=0,1,2$. The representation $\RR_I$
associates to $\theta$ the phase $e^{2\pi i I/3}$. Clearly 
the product law in the set of irreducible representations is 
$\RR_{I}\otimes 
\RR_{J}=\RR_{I+J}$, i.e. amounts to usual addition (modulo 3) of 
subindices. We see from (\ref{actionz3}) that the 
representation ${\bf 3}$ that we have chosen decomposes as $\RR_1\oplus 
\RR_1\oplus \RR_1$. 

Following the rules above, the gauge group is $SU(N)_0 \times SU(N)_1 
\times SU(N)_2$. The fields of type $\Phi_{I,I+A_1}$, associated to the 
first complex plane, transform as a copy of
$(\fund_{0},\antifund_{1})$$ + (\fund_{1},\antifund_{2})$$ + 
(\fund_{2},\antifund_{3})$. We will denote these fields as $Q_I^1$. The fields 
$\Phi_{I,I+A_2}$, which we denote by $Q_I^2$, are associated to the second 
complex plane, and transform as another copy of the same representation. 
Finally, the fields $\Phi_{I,I+A_3}$ transform again in this representation. 
We denote them by $Q_I^3$. So in 
total the model has nine chiral multiplets $Q_I^a$ transforming in the 
representation
\beq
3 \;(\fund_{0},\antifund_{1})\; +\;  3\;(\fund_{1},\antifund_{2})\; +\; 
3\; (\fund_{2},\antifund_{3}).
\eeq
Following the rules 
above, there is a superpotential which can be rewritten as
\beq
W\sim \epsilon^{IJK} Q_I^1 Q_J^2 Q_K^3
\eeq

This model was studied in \cite{ibanez,ks} before the 
general formulation of \cite{lnv} appeared.

\medskip

{\bf ii) $C^3/(Z_k \times Z_{k'})$}

Let us consider a rather large family of singularities of type 
$\IC^3/(\IZ_k\times\IZ_{k'})$, which will be useful in the following 
sections. Let $\theta$, $\omega$ denote the 
generators of the $\IZ_k$, $\IZ_{k'}$ subgroups, respectively. The group 
$\Gamma=\IZ_k\times \IZ_{k'}$ has $kk'$ irreducible representations, 
denoted $\RR_{a,b}$ 
($a=0,\ldots,k-1$, $b=0,\ldots,k'-1$). The representation $\RR_{a,b}$ 
associates to the general element $\theta^m\omega^n$ the phase factor 
$e^{2\pi i \frac{am}{k}} e^{2\pi i\frac{bn}{k'}}$. Notice that {\em one} 
uppercase index in the general formulation represents {\em two} 
lowercase indices 
in this case, since the group has two generators. The product of 
representations is given by $\RR_{a,b}\otimes \RR_{a',b'} = 
\RR_{a+a',b+b'}$, {\em i.e.} separate addition in the indices.

Let us choose the action of $\Gamma$ on $\IC^3$ such that the generators 
act as
\beqa
\theta \; : & (z_1,z_2,z_3) \to & (e^{2\pi i/k} z_1,z_2,e^{-2\pi i/k} z_3)
\nonumber \\
\omega \; : & (z_1,z_2,z_3) \to & (z_1,e^{2\pi i/k'} z_2,e^{-2\pi 
i/k'}z_3)
\label{actionzkzkp}
\eeqa
This means that the corresponding representation ${\bf 3}$ decomposes as 
${\bf 3} = \RR_{1,0}\oplus \RR_{0,1}\oplus \RR_{-1,-1}$. 

The gauge group is $\prod_{a,b} SU(N)_{a,b}\approx SU(N)^{kk'}$. The 
chiral multiplets associated to the first complex plane, $\Phi_{I,I+A_1}$, 
form the representation $\bigoplus_{a,b} (N_{a,b},{\ov N_{a+1,b}})$. 
Similarly, the fields $\Phi_{I,I+A_2}$ are in the representation 
$\bigoplus_{a,b} (N_{a,b},{\ov N_{a,b+1}})$, and the fields 
$\Phi_{I,I+A_3}$ transform as $\bigoplus_{a,b} (N_{a,b}, {\ov 
N_{a-1,b-1}})$.

The resulting spectra are rather lengthy to list, but straightforward to 
obtain. Similarly, the superpotential terms follow from equation 
(\ref{superp}).

We postpone the discussion of how the field theory parameters are encoded 
in the configuration of branes at singularities until Section~5. In the 
meantime, in sections~3 and 4, we establish a connection between the 
brane box models and the branes at singularities. This relation will allow 
us to map several parameters, states, and field theory phenomena from the 
brane box configurations to the singularity language.

\section{From the singularity to the brane box}

\subsection{The construction}

The general pattern of the theories that we have obtained studying D3 
branes on top of $\IC^3/\Gamma$ singularities, for Abelian $\Gamma$, is 
very reminiscent of the type of theories we obtained using the brane box 
configurations. Our aim in this section is to show that actually for each 
such singularity theory one can construct a suitable brane box 
configuration leading to the same four-dimensional $\NN=1$ gauge theory. 
We stress that the argument is simply based on the matching of the 
spectra, and does not establish a physical principle underlying the 
correspondence. We will come back to this point in the following section, 
where we show the relation follows from T duality.

\begin{figure}
\centering   
\epsfxsize=6in
\hspace*{0in}\vspace*{.2in}
\epsffile{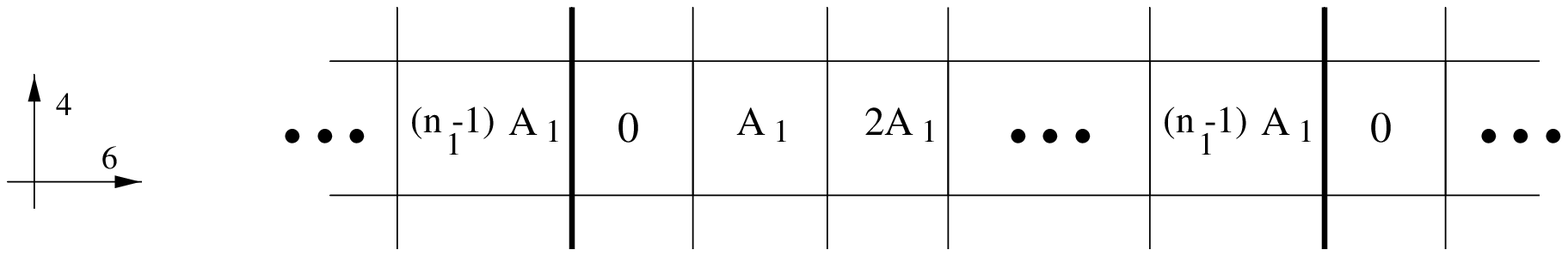}
\caption{\small First step in the construction of the brane box model 
corresponding to a singularity $\IC^3/\Gamma$. The labels on each box 
denote the irreducible representation of $\Gamma$ associated to the gauge 
factor in that box. The basic period in the infinite array of boxes is 
bounded by dark lines.} 
\label{fig:step1}
\end{figure}

The general strategy to construct such a brane box configuration is to 
draw one box for each irreducible representation of $\Gamma$, so as to 
ensure the gauge group 
is the same, and adjoin these boxes such that the chiral multiplets 
H, V, D  in the box model reproduce the 
fields $\Phi_{I,I\oplus A_1}$, $\Phi_{I,I\oplus A_2}$ and $\Phi_{I,I\oplus 
A_3}$. Notice that in the brane box diagram the relation of 
neighbourhood of boxes will thus be defined in terms of the product law of 
irreducible representations in $\Gamma$. The construction of the brane 
boxes is thus very similar to that of the quiver 
diagrams described in \cite{reid}.

The construction is easily organized as follows: The first step is to draw 
a row of boxes corresponding to the irreducible representation $\RR_0$, 
$\RR_{A_1}$, 
$\RR_{2 A_1}$, $\ldots$, $\RR_{(n_1-1)A_1}$, where $n_1$ is the order of 
$\RR_{A_1}$ in the set of irreducible representations of $\Gamma$. The 
fact that $\RR_{n_1 A_1}\equiv 
\RR_0$ means that the row is compactified on a circle. Equivalently, one 
can think of a one-dimensional infinite periodic array of boxes, of which 
the finite set described above is a fundamental region. This construction 
ensures that the horizontal fields between the boxes reproduce some of the 
fields $\Phi_{I,I+A_1}$. A picture of this first step 
in the construction is shown in figure \ref{fig:step1}.

\begin{figure}
\centering   
\epsfxsize=5in
\hspace*{0in}\vspace*{.2in}
\epsffile{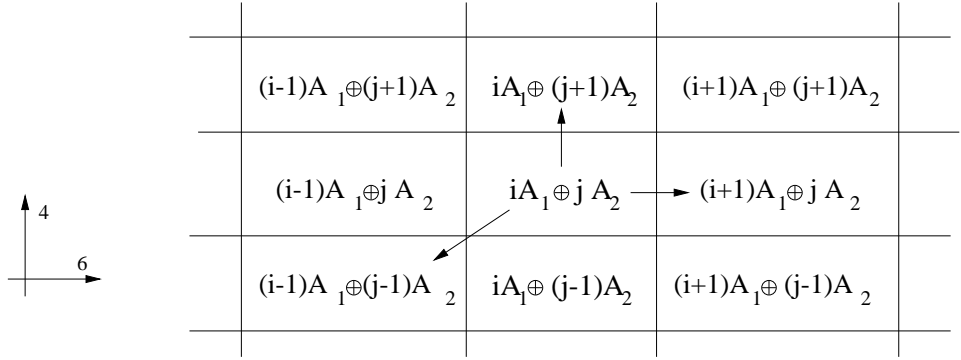}
\caption{\small A typical region in a brane box model constructed from a 
singularity. The arrows denote the chiral multiplets $\Phi_{I,I\oplus 
A_1}$, $\Phi_{I,I\oplus A_2}$, and $\Phi_{I,I-A_1-A_2}$, for 
$I=iA_1\oplus jA_2$, which appear as the horizontal, vertical and diagonal 
fields.} 
\label{fig:step2}
\end{figure}

Next, from each of the boxes $\RR_{nA_1}$ in the row, we start a vertical 
column of boxes, 
which we label $\RR_{nA_1}$, $\RR_{nA_1\oplus A_2}$, $\ldots$, 
$\RR_{nA_1\oplus (n_2-1)A_2}$, where $n_2$ is the order of $\RR_{A_2}$ in 
the set of irreducible representations. Again, since $\RR_{n_2 A_2}\equiv 
\RR_0$, there is 
an identification of the horizontal sides of the resulting rectangle, 
which makes the configuration to be compactified on a two-torus. 
Alternatively, one can extend the construction to the full plane, and 
think of it as the universal cover of the torus. This construction ensures 
that the new horizontal arrows reproduce the fields of type 
$\Phi_{I,I\oplus A_1}$, and the vertical arrows the multiplets of type 
$\Phi_{I,I\oplus A_2}$. Quite remarkably, the fields of type 
$\Phi_{I,I+A_3}$ are automatically reproduced by the diagonal arrows, and 
the superpotential interactions (\ref{superp}) are reproduced by the 
triangle rule (\ref{HVD}). In figure \ref{fig:step2} we show a 
typical region in a general brane box thus constructed.
As can be read in the picture, when one moves horizontally to the right, 
the label in the boxes shifts by $A_1$; when one moves vertically upwards, 
the label changes by $A_2$; finally, a diagonal movement from upper-right 
to lower-left shifts the label by $-A_1-A_2$.

An important question is whether all irreducible representations of 
$\Gamma$ indeed
appear in this rectangle. That this is so follows from the fact that 
the representation ${\bf 3}$ was chosen to be faithful. This means that 
all irreducible representations of $\Gamma$ are generated by $\RR_{A_1}$, 
$\RR_{A_2}$.
As an example of a non-faithful representation, consider the case where 
$\Gamma=\IZ_8$ and the representation
{\bf 3} decomposes as $\RR_2\oplus\RR_2\oplus \RR_4$.
It is easy to construct the brane configuration and to discover that it
actually describes the case with $\Gamma=\IZ_4$ and the representation {\bf 3}
decomposes as $\RR_1\oplus\RR_1\oplus \RR_2$.

\begin{figure}
\centering   
\epsfxsize=5in
\hspace*{0in}\vspace*{.2in}
\epsffile{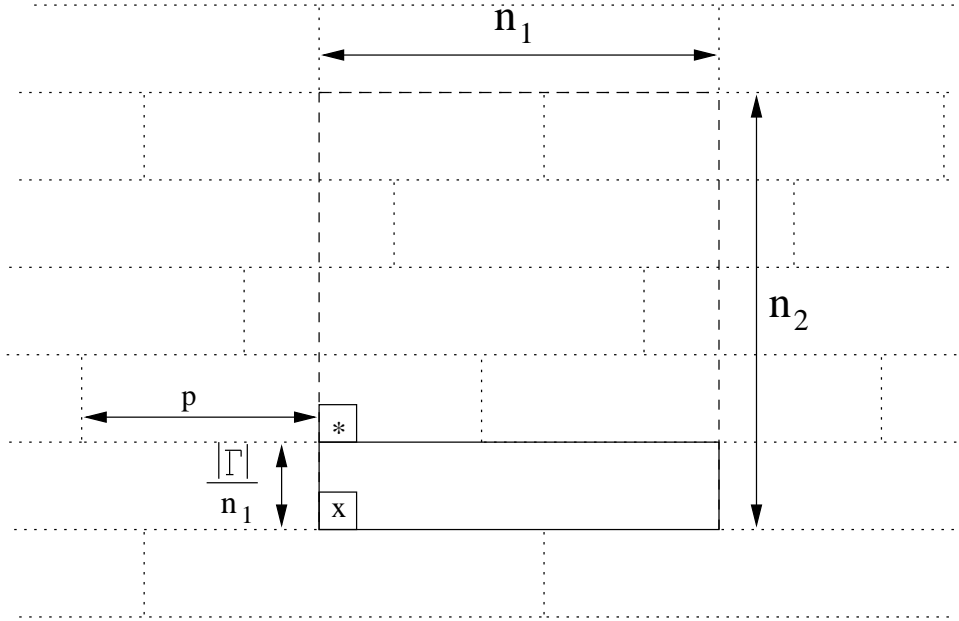}
\caption{\small The relation between the $n_1\times n_2$ box rectangle and 
the true unit cell in a general brane box configuration.}
\label{fig:unitcell}
\end{figure}

Another 
related point is whether all the $n_1 n_2$ boxes are actually different. 
In 
general, this is not so, and each box will be repeated $q=n_1 
n_2/|\Gamma|$ 
times. Since this $n_1 \times n_2$ box region is the {\em smallest} 
rectangle with {\em trivial} identifications of sides, and the true unit 
cell (where each box appears once and only once) is {\em smaller}, it will 
have {\em non-trivial} identification of sides. The true unit cell can be 
obtained as 
a $n_1 \times |\Gamma|/n_1$ cell in the rectangle. The relation between 
the $n_1\times n_2$ rectangle and the unit cell is illustrated in figure 
\ref{fig:unitcell}.  It is clear that the 
identification of vertical sides of the unit cell will be trivial. 
However, the identifications of horizontal sides is accompanied by a 
shift. If the box marked with a `x' in the picture is labeled $\RR_{0}$, 
the box marked with a `*' is labeled $\RR_{(|\Gamma|/n_1)A_2}$, which 
is equal 
to $\RR_{pA_1}$ for some integer $p$. The identification of horizontal 
sides is shifted by $p$ boxes to the left.

\begin{figure}
\centering   
\epsfxsize=3.5in
\hspace*{0in}\vspace*{.2in}
\epsffile{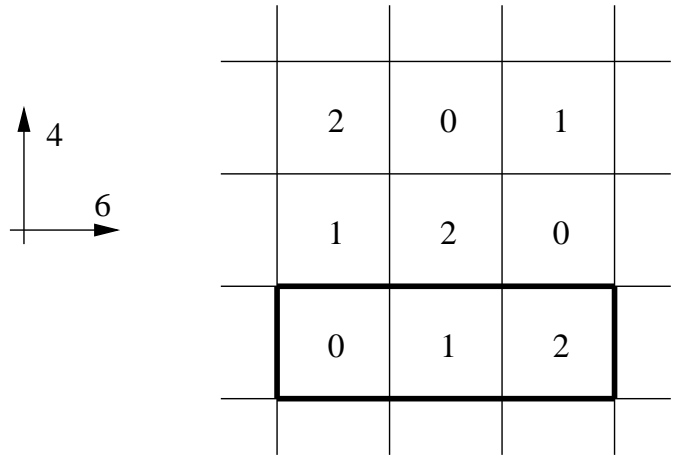}
\caption{\small The brane box corresponding to a $\IC^3/\IZ_3$ 
singularity. It is 
a $3\times 1$ box model with trivial identifications of vertical sides, 
and identification of horizontal sides up to a shift of one box.}
\label{fig:z3}
\end{figure}

As an illustrative case consider the orbifold $\IC^3/\IZ_3$, discussed as 
example i) in section~2.2, where we had ${\bf 
3}=\RR_1\oplus\RR_1\oplus\RR_1$. Since the order of $\RR_1$ in the set of 
irreducible representations is 3, we have $n_1=3$ and also $n_2=3$. The 
procedure we have described yields a $3\times 3$ rectangle with trivial 
identifications, which is depicted in figure \ref{fig:z3}. Each box is 
repeated 3 times ($q=3$), so the true unit cell is smaller, and has 
non-trivial identifications. The unit cell can be taken to be the 
$3\times 1$ cell highlighted in the figure. The vertical sides are 
identified in the trivial way, but the horizontal identification is 
accompanied by a shift of one box ($p=1$). This brane configuration was 
introduced in \cite{hsu}, where it was already observed that its spectrum 
matched that of D3 branes at a $\IZ_3$ singularity.

\medskip

An important point is that the consistency of both constructions is only 
possible for {\em Abelian} discrete groups $\Gamma$. This is suggested 
from a number of perspectives. For example, if one wishes to construct 
{\em finite} theories from D3-branes sitting at singularities, one should 
choose the Chan-Paton embedding as $N$ copies of $\RR_{\Gamma}$. The gauge 
group for a general $\Gamma$ is $\prod_I SU(N n_I)$, where $n_I$ is the 
dimension of the $I^{th}$ irreducible representation. In the construction 
of {\em finite} theories using brane box configurations \cite{hsu}, the 
gauge group is $SU(N)^{M}$, where $M$ is the number of boxes. Thus the 
brane box configurations reproduce some of the finite models that can be 
constructed from singularities, namely those where $\Gamma$ is Abelian and 
all the irreducible representations are one-dimensional, $n_I=1$, $\forall 
I$.

Also, when $\Gamma$ is Abelian the tensor product of representations is 
commutative, so $\RR_{I\oplus A_1\oplus A_2}\equiv\RR_{I\oplus A_2\oplus 
A_1}$. This is a necessary requirement in our construction of the brane 
boxes, since it ensures that, starting from the box labeled by $\RR_I$ and 
moving one box to the right and then one box upwards one reaches the same 
box than moving first upwards and then to the right, an unavoidable 
geometrical fact in the brane box construction.

This restriction on the type of singularity is hardly a surprise. Several 
works in geometric engineering (see {e.g.} \cite{vafa}) have shown that the 
geometric 
approach to the realization of gauge theories is more {\em general} than 
the constructions using brane configurations. However we would like to stress 
that in many instances the brane configurations provide a {\em simpler} 
realization of the gauge theories. In our particular case, we are to see 
that configurations which are {\em not} finite are easily constructed and 
analyzed in the brane configuration language, by simply putting a 
different number of D5 branes in each box. Reproducing these theories 
using branes at singularities requires the use of fractional branes, 
objects which are forced to lie at the singular point. Since 
basically everything is happening at the singular point, the construction 
is much less intuitive. Somehow, the brane  picture `opens up' all the
phenomena happening at the singularity and displays them in different 
boxes.

\medskip

As a final comment, let us mention that there is an arbitrary choice in 
the procedure above, namely the different ways to assign the three 
kinds of fields H, V, D to the three kinds of fields $\Phi_{I,I+A_i}$ 
$i=1,2,3$. There are six such inequivalent choices. The brane box 
configurations one obtains are related in such a way that, e.g. 
the fields that arise from horizontal arrows in one come from diagonal 
arrows in the other. This is clearly the transformation of brane box 
models we introduced in 
Section~2. The meaning of this transformation will become clear 
after the T-duality relation between branes boxes and branes 
at singularities is established in the next section.

\medskip

\subsection{Examples}

To illustrate the ideas we have introduced, we present a few examples of 
the construction above.

\begin{figure}
\centering   
\epsfxsize=6in
\hspace*{0in}\vspace*{.2in}
\epsffile{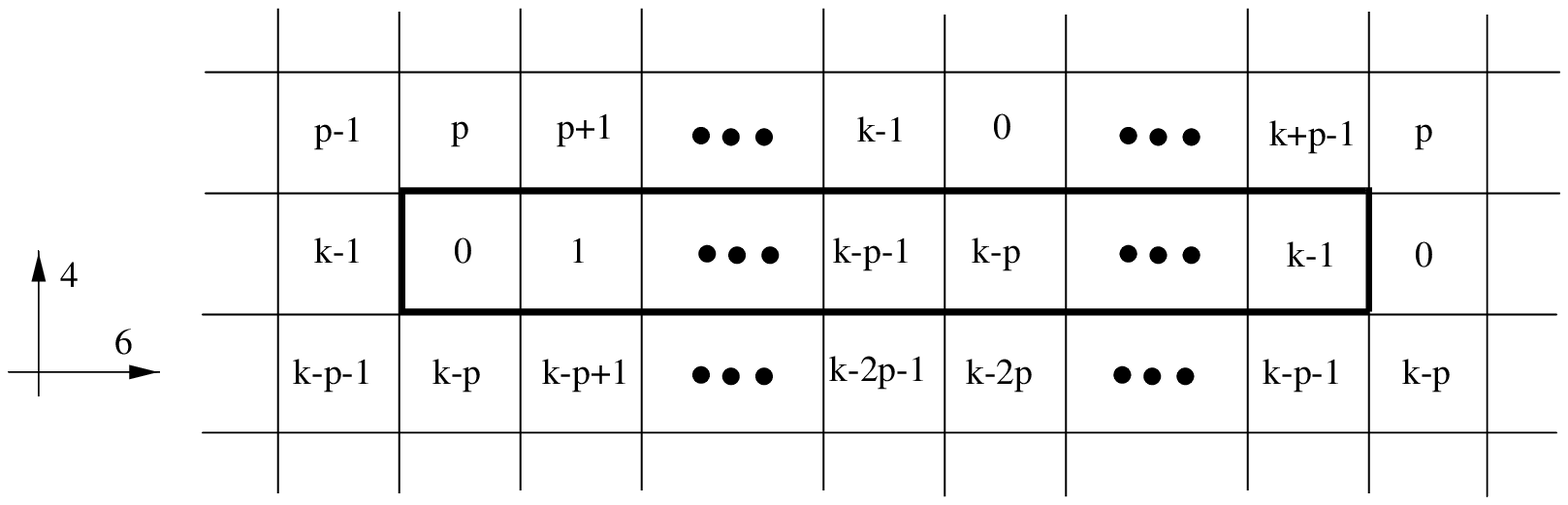}
\caption{\small The brane box configuration obtained form the 
$\IC^3/\IZ_k$ 
singularity described in the text. It corresponds to a $k\times 1$ box 
model with trivial identifications of vertical sides and 
identifications of the horizontal sides up to a shift of $p$ boxes to 
the left.} 
\label{fig:zkshifted}
\end{figure}

{\bf i)} Consider a singularity $\IC^3/\IZ_k$, with the generator $\theta$ 
of $\IZ_k$ acting on $\IC^3$ as
\beq
\theta \; : \; (z_1,z_2,z_3) \to (e^{2\pi i/k} z_1,e^{2\pi i \frac{p}{k}} 
z_2, e^{2\pi i \frac{(-p-1)}{k}} z_3)
\eeq
with $p$ an integer in the range $0\leq p \leq k-1$.
The representation ${\bf 3}$ we have chosen decomposes as 
$\RR_{1} \oplus \RR_{p}\oplus\RR_{-p-1}$. It is easy to check that the 
above procedure yields the brane box diagram shown in figure 
\ref{fig:zkshifted}. 
Starting from any box, a horizontal movement to the right shifts the label 
by $1$, so that the horizontal arrows give rise to the fields 
$\Phi_{I,I+1}$. A vertical movement upwards shifts the label by $p$, so 
that vertical arrows correspond to $\Phi_{I,I+p}$. And a diagonal 
movement downwards and to the left shifts the label by $-p-1$, so that
diagonal fields correspond to $\Phi_{I,I-p-1}$. 

The order of $\RR_{A_1}$ is $k$. For $p\neq 0$ the order 
of $\RR_{A_2}$ is $k/\ell$, where $\ell$ is the greatest common divisor of 
$k$ and $p$, $\ell=\gcd(k,p)$. Our procedure above yields a rectangle of 
$k\times k/\ell$ boxes with trivial identifications. Each box is repeated 
$k/\ell$ times, so the unit cell is smaller and has non-trivial 
identification of sides. In figure \ref{fig:zkshifted} we show a choice of 
unit cell, 
which has trivial identifications of vertical sides and the horizontal 
identification up to a shift of $p$ boxes to the left.
For $k=3$, $p=1$, we recover the $\IZ_3$ example previously studied.

\medskip

\begin{figure}
\centering   
\epsfxsize=5in
\hspace*{0in}\vspace*{.2in}
\epsffile{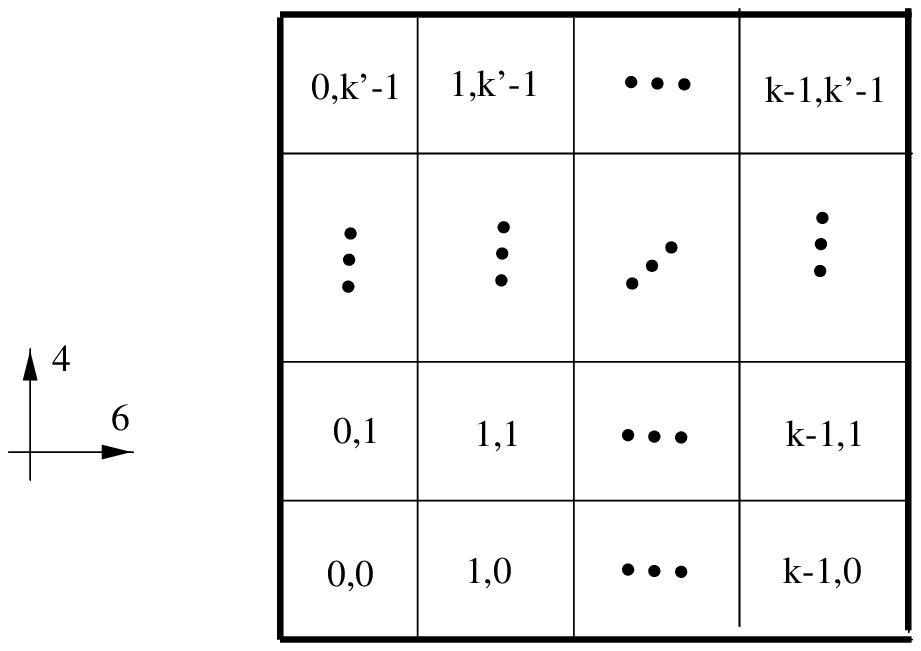}
\caption{\small The unit cell of a $k\times k'$ box model with trivial 
identifications, as obtained from a $\IC^3/(\IZ_k\times \IZ_{k'})$ 
singularity.}
\label{fig:zkzkprime}
\end{figure}

{\bf ii} As a last example consider the singularities of type 
$\IC^3/(\IZ_k\times 
\IZ_{k'})$, with the action on $\IC^3$ defined as in (\ref{actionzkzkp}), 
namely ${\bf 3} = \RR_{1,0}\bigoplus\RR_{0,1}\bigoplus\RR_{-1,-1}$.
It is easy to realize that the spectra of the gauge theories on D3 branes 
on these singularities can be obtained by D5 branes on a grid of $k\times 
k'$ boxes, with trivial identifications of sides. One such grid is shown 
in figure \ref{fig:zkzkprime}, where each box is labeled by its associated 
irreducible representation. 
Horizontal movements to the right change 
the label by multiplication by $\RR_{1,0}$, vertical movements upwards 
correspond to multiplication by $\RR_{0,1}$, and diagonal movements to 
multiplication by $\RR_{-1,-1}$. Since the order $n_1$ of $\RR_{1,0}$ in 
the set of irreducible representations is $k$, and the order $n_2$ or 
$\RR_{0,1}$ is $k'$ in this case the $n_1\times n_2$ rectangle coincides 
with the unit cell.

\section{T-duality: from the brane box to the singularity}

\subsection{Some simple examples}

In this section we will explain the reason underlying the precise matching 
found above between the spectra of brane box configurations and D3 branes 
on singularities. Specifically, we argue that the relation is a T-duality 
between both kinds of constructions. We will show how, starting from a 
brane box configuration and T-dualizing along 4 and 6, the geometry around 
the D3 branes in the dual is locally $\IC^3/\Gamma$, with $\Gamma$ an 
abelian subgroup of $SU(3)$. We also provide an explicit construction of 
the $\Gamma$ corresponding to a given brane box configuration.

The D5 branes will not play any relevant role in the T-duality 
relation between the NS and NS' branes and the singularity. For 
simplicity it is convenient to consider the case in which the number of D5 
branes in each box is the same. More general configurations will be 
analyzed in Section~6.

The T-duality transformation is quite analogous to that relating a set of 
$k$ parallel NS branes and an $A_{k-1}$ ALE space, so it is convenient to 
briefly review some of its features. Consider a Type IIB configuration of 
$k$ parallel NS branes extending along 012345, and let $x^4$, $x^6$ be 
compact coordinates. If $N$ D5 branes are located along 012346 (wrapping 
the two-torus in 46) the configuration provides a realization of the 
$SU(N)^k $ $\NN=2$ elliptic models \cite{wit4d} on the D brane 
worldvolume. The T-duality 
considerations in this case have already been explored in \cite{lust}. 

Performing a T-duality along 4,6, the D5 branes are mapped to D3 branes 
along 0123, sitting at a point in the T-dual coordinates, which we denote 
by 4$'$, 6$'$. The duality along 4 is longitudinal to the $k$ NS branes, 
and does not change them, while the duality along 6 transforms them into 
$k$ Kaluza-Klein (KK) monopoles. Thus, the space parametrized by 6$'$789 
in the 
T-dual is a $k$-centered multi-Taub-NUT space, described by the metric
\beqa
ds^2\; & = & \; \frac {V}{4} d{\vec r\,}^2 \, +\, 
\frac{V^{-1}}{4}(dx^{6'}+{\vec \omega}\cdot d{\vec r})^2, \nonumber \\
{\rm with} & & V\; =\; 1+ \sum_{i=1}^k \frac{1}{|{\vec r}-{\vec x}_i|}
\label{taubnut}
\eeqa
and ${\vec \nabla}\times \omega={\vec \nabla} V$. This is a fibration of
an $S^1$ (parametrized by $x^{6'}$) over $\IR^3$ (parametrized by ${\vec 
r}=(x^7,x^8,x^9)$), the fibers of which shrink to zero radius at the $k$ 
centers ${\vec x_i}$. The parameters in the original theory can be traced 
to the final configuration. For example, the position of the $k$ NS 
branes on 789 are mapped to the positions of the $k$ centers ${\vec x_i}$. 
An interesting remark in this respect is that, when all such positions 
coincide in the brane box configuration, all the centers in the Taub-NUT 
space coalesce at a point. For ${\vec r}$ very close to this point, the 
constant term in 
$V$ in equation (\ref{taubnut}) can be neglected, and the geometry is that 
of an ALE singularity. If in the initial picture the D5 branes also sit at 
$x^7=x^8=x^9=0$, the D3 branes will be located at the singular point, and 
the physics of the gauge theory on their worldvolume is controlled by the 
structure of the $A_{k-1}$ singularity. This provides the connection with 
the 
description in \cite{dm,lnv}. When 
the positions of the centers differ from each other, the 
singularity is resolved and the number of factors in the gauge theory is 
reduced, by Higgs breaking to diagonal subgroups \footnote{Notice that 
here we are considering blow-ups of small size as compared with the 
Taub-NUT radius, so that the approximation of the metric as an ALE space 
remains valid.}. This is the same 
breaking that occurs in the initial brane box configuration when the 
positions of the NS branes are slightly changed.

It may seem that the positions of the NS branes on $x^6$ have been lost in the 
T-duality. However, as shown for instance in \cite{sen}, they are actually 
encoded 
in the singularity picture as integrals of the NS-NS two-form $B_{NS}$ 
over the non-trivial 2-cycles of the Taub-NUT (or ALE) space. Such 
two-spheres, which we will denote by $\Sigma_{ij}$ are 
obtained as the fibration of the $S^1$ parametrized by $x^{6'}$ over the 
segments joining the centers ${\vec x_i}$ and ${\vec x_j}$. A basis of 
$k-1$ two-cycles is provided by 
$\Sigma_{i,i+1}$, for $i=1,\ldots,k-1$. So, the $k-1$ independent 
quantities
\beq
a_i\; =\; \int_{\Sigma_{i,i+1}} B_{NS}.
\label{flux1}
\eeq
provide the $k-1$ independent distances between NS branes (more 
precisely, they provide the ratios of such distances to the total length 
$R_6$ of the $x^6$ direction).
There is also a set of analogous parameters corresponding to 
\beq
v_i\; =\; \int_{\Sigma_{i,i+1}} B_{RR}
\label{flux2}
\eeq
where one is integrating the Ramond-Ramond two-form field over the basic 
two-cycles. As discussed in \cite{hsu}, these correspond to the 
differences of Wilson lines along $x^4$ of the world-volume $U(1)$ gauge 
fields of the original $k$ NS branes (more precisely, the ratios of such 
differences to dual radius $1/R_4$). The parameters 
$a_i$, $v_i$ define the gauge couplings of the $\NN=2$ gauge theory 
\cite{lnv,hsu}, as also follows by particularizing equations 
(\ref{gauge}), (\ref{theta}) to this $\NN=2$ case.

The Coulomb branch of the gauge theory in the singularity 
language is parametrized by movements of the D3 branes in 4$'$5 keeping 
the coordinates in 6$'$789 fixed at the singularity. As we have mentioned, 
there are Higgs branches which correspond to resolving partially the 
singularity (these map to the removal of the corresponding NS brane in 
the brane box picture). Finally, there is also a Higgs branch 
corresponding to moving the D3 branes away from the singularity (this 
branch maps to 
recombining the D5 branes and moving them away from the grid of NS and 
NS' branes).

Thus we see how all the information of the brane box configuration is 
encoded in the singularity, and vice-versa. Since there are aspects of the 
gauge theory which are easier to analyze in either of both pictures, we 
hope the dictionary we intend to develop in the present work will also be 
useful in the understanding of general chiral $N=1$ gauge theories.

\medskip

The next example we would like to consider is a $k \times k'$ box model, 
with trivial identifications of the sides of the unit cell, as that shown 
in 
figure \ref{fig:zkzkprime}. Thus we start with $k$ NS branes along 012345 
and $k'$ NS$'$ branes along 012367. We will consider the case of 
having an equal number $N$ of D5 branes with world-volume filling 012346.
After a 
T-duality along 4 and 6, the $k$ NS branes are transformed into $k$ KK 
monopoles, realized as a non-trivial geometry in the coordinates 6$'$789. 
On the other hand the $k'$ NS$'$ branes become KK$'$ monopoles, 
corresponding to a nontrivial geometry in 4$'$589. The resulting space in 
4$'$,5,6$'$,7,8,9 is a non-compact Calabi-Yau threefold for which we do 
not 
have an explicit metric. However, since it arises by the `superposition' of 
multi-Taub-NUT metrics, it is easy to uncover the relevant features which 
control the gauge theory on the D3 brane probes. If we consider the regime 
where the positions on 789 of the NS branes are close to the origin in the 
initial configuration, it is clear that the centers of the KK monopoles will 
lie close to the origin. The space contains a curve of $A_{k-1}$ ALE
singularities parametrized by 4$'$5 and roughly defined by 
$x^7=x^8=x^9=0$. Similarly, when the positions of the NS$'$ branes on 589 
are close to the origin, the space will contain a curve of $A_{k'-1}$ ALE 
singularities defined by $x^5=x^8=x^9=0$, and parametrized by 6$'$7. Both 
curves intersect at a point, where the 
singularity is worse and has the local structure of $\IC^3/(\IZ_k\times 
\IZ_{k'})$, with the generators $\theta,\omega$ of $\IZ_k$, $\IZ_{k'}$ 
acting on $(z_1,z_2,z_3)\in \IC^3$ as 
\beqa
\theta: & (z_1,z_2,z_3) & \to (e^{2\pi i/k} z_1,z_2,e^{-2\pi i/k} z_3) 
\nonumber \\
\omega: & (z_1,z_2,z_3) & \to (z_1,e^{2\pi i/k'} z_2,e^{-2\pi i/k'} z_3) 
\eeqa
Here the complex coordinate $z_1$ corresponds to $x^7, x^{6'}$, the 
coordinate $z_2$ refers to $x^5, x^{4'}$, and $z_3$ to $x^8,x^9$. 

The D3 branes will sit precisely at the singular point, so the structure 
of the 
singularity controls the properties of the $\NN=1$ four-dimensional gauge 
theory. This T-duality argument explains the result we obtained in 
Section~3 where we observed that the theory on D3 branes on top of such 
singularity was the same as that obtained in a $k\times k'$ box model.

Finally, and for future convenience, let us notice that through this 
T-duality map, the box located in the position $(i,j)$ in the $k\times k'$ 
box grid corresponds to the irreducible representation $\RR_{i,j}$ of 
$\IZ_k\times \IZ_{k'}$. This is manifest from our example ii) of 
section~3, and will be a useful way of labeling boxes in some arguments.

\subsection{Models with non-trivial identifications}

The only difficulty in extending the above arguments to a T-duality 
prescription for a general brane box configuration is the possibility of 
identifications up to a shift.
In the following we show how to handle these cases. Consider a general 
brane box model, which without loss of generality, we can take to 
be a $k\times k'$ box model with trivial identification of the vertical 
sides, and identification of horizontal sides with a shift of $p$ boxes to 
the left. Our aim is to use T-duality to 
relate this configuration to some geometry (which admits a local 
description as $\IC^3/\Gamma$) such that the gauge theory on D3 brane 
probes reproduces the initial one.

\begin{figure}
\centering   
\epsfxsize=4in
\hspace*{0in}\vspace*{.2in}
\epsffile{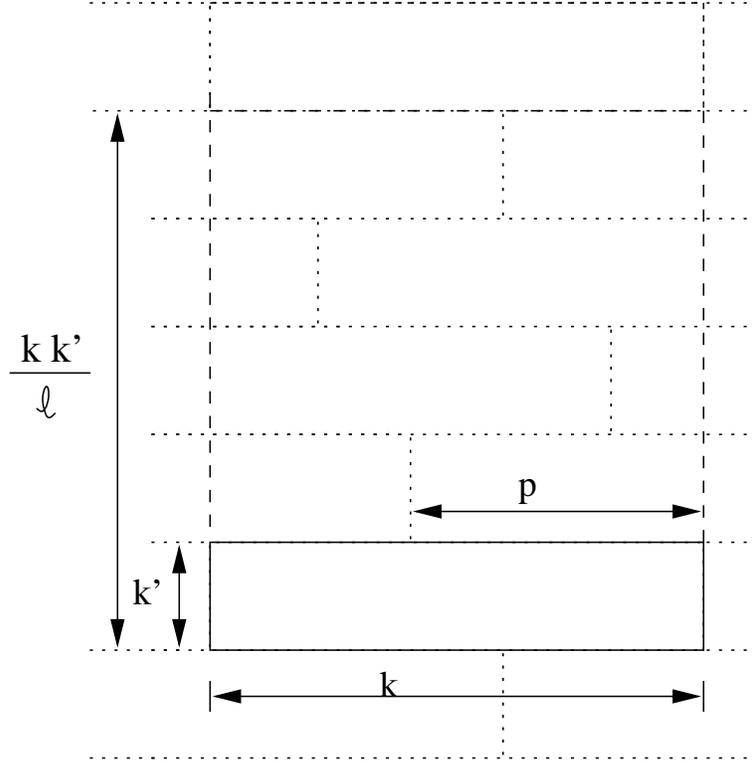}
\caption{\small By adjoining several unit cells, we can define a larger 
rectangle whose sides have trivial identifications.}
\label{fig:extend}
\end{figure}

In order to avoid the complications coming from the shifted 
identification, we can adjoin several unit cells until we fill a rectangle 
for which the identifications of sides are trivial \footnote{This 
process is the inverse of that in Section~3, where starting from a 
rectangle with trivial identifications we made a choice for the unit 
cell inside it.}. Figure \ref{fig:extend} shows  
a way of doing this. If we denote by $\ell$ the greatest common divisor of 
$k$ and $p$, $\ell=\gcd(k,p)$, such a rectangle has $k \times kk'/\ell$ 
boxes. 
On it, the true unit cell is repeated $k/\ell$ times. However, it will be 
useful to consider a `parent' model where all the $k \times kk'/\ell$ 
boxes are considered
independent. Our original model will be obtained from this one after a 
$\IZ_{k/\ell}$ identification, given by a translation in the torus by $p$ 
boxes to the left and $k'$ boxes upwards.

The strategy we are to follow is first to T-dualize along 46, and then to 
impose this identification in the resulting T-dual picture. The T-duality 
of the parent model presents no difficulty since the sides of the 
rectangle have trivial identifications. The T-dual theory 
is that of D3 branes sitting at a singularity locally of the form 
$\IC^3/(\IZ_k\times \IZ_{kk'/\ell})$, with the generators $\theta$, $\omega$ 
of $\IZ_k$, $\IZ_{kk'/\ell}$ acting as
\begin{eqnarray}
(z_1,z_2,z_3) & \to & (e^{2\pi i/k}z_1,z_2,e^{-2\pi i/k}z_3) 
\nonumber \\
(z_1,z_2,z_3) & \to & (z_1,e^{2\pi i\frac{\ell}{kk'}}z_2,e^{-2\pi 
i\frac{\ell}{kk'}}z_3) 
\label{action}
\end{eqnarray}
with $z_1,z_2,z_3$ defined as in the preceding subsection.

Since our original model had fewer different boxes  than the parent 
model, the final theory should correspond to D3 branes sitting at a less 
singular point. In order to understand how this can be done, it is 
illuminating to momentarily consider a similar problem in a $\NN=2$ 
theory. Consider such theory realized as a $k\times 1$ box model with 
trivial identifications of sides, and perform a T-duality along 46. This 
yields, as we know, D3 branes at a $\IC^2/\IZ_k$ singularity with 
generator $\theta$ acting as $(z_1,z_3)\to (e^{2\pi i/k} z_1,e^{-2\pi i/k} 
z_3)$ 
(forgetting about $x^4,x^5$ which does not enter the argument in this 
simpler case). But let us suppose we make a `mistake' in the choice of the 
unit cell and consider it to be a $nk\times 1$ box rectangle, without 
noticing that each box is repeated $n$ times. So, if the $nk$ boxes are 
considered different, we end up with a T-dual geometry $\IC^2/\IZ_{nk}$, 
with generator $\theta'$ acting as $(z_1,z_3)\to (e^{2\pi i 
\frac{1}{nk}} z_1,e^{-2\pi i\frac{1}{nk}} z_3)$. The question is how we 
can correct our `mistake', the $\IZ_n$ identification we had missed, once 
in the dual picture. This is done by noticing that the true T-dual should 
correspond to $\IC^2/\Gamma$ with $\Gamma$ a subgroup of $\IZ_{nk}$. In 
this case we have $\Gamma=\IZ_{nk}/\IZ_n$$\approx \IZ_k$ with generator 
$\theta=\theta'^n$. This $\IZ_n$ action can be viewed as an order $n$ 
automorphism on the extended Dynkin diagram of $A_{nk-1}$, a 
counter-clockwise rotation by $n$ nodes. This action is actually 
geometrically realized on the two-cycles that resolve the singularity.

A similar discussion applies to our $\NN=1$ case. We had obtained a 
$\IC^3/(\IZ_{k}\times \IZ_{kk'/\ell})$ singularity, but the parent model 
contained an order $k/\ell$ identification of boxes, which we had not 
taken 
into account. The true singularity then must correspond to 
$\Gamma=$$(\IZ_{k}\times \IZ_{kk'/\ell})/\IZ_{k/\ell}$. The $\IZ_{k/\ell}$ 
is 
generated by $\theta^{-p}\omega^{k'}$ (as can be seen by noticing the 
action relating two identified boxes in the original picture). The 
representation ${\bf 3}$ that defines the action of $\Gamma$ on $\IC^3$ is 
induced from the action of the parent singularity, in the following way. 
One defines a surjective homomorphism from the set of irreducible 
representations of the parent 
singularity to the set of irreducible representations ${\bf \RR_I}$ of 
$\Gamma$, such that 
$\RR_{-p,k'}$ is mapped to the unity ${\bf \RR_0}$. Then $\Gamma$ acts on 
$\IC^3$ as dictated by the image of ${\bf 3}$ through this homomorphism. 
The resulting singularity is independent of the particular homomorphism 
chosen, as long as it fulfills the mentioned condition.
\medskip

The meaning of the above procedure is most clear when we recall from 
previous sections that each box in the brane box configuration corresponds 
to one irreducible representation of $\Gamma$. The fact that some boxes in 
the $k\times kk'/\ell$ rectangle are identical means that some 
representations in $\IZ_k\times \IZ_{kk'/\ell}$ are to be considered 
identical. This is accomplished by the homomorphism above, which 
essentially states that a movement of $p$ boxes to the left and $k'$ 
upwards takes one box to another copy of the same box. 

\medskip

{\large \bf Examples}

To make the construction somewhat clearer, let us work out a few 
examples. 

{\bf i)} Let us start considering the $3\times 1$ box model with trivial 
vertical identifications, and horizontal identifications up to a shift of 
one box to the left, as shown in figure \ref{fig:z3}. Thus $k=3$, $k'=1$, 
$p=1$ and $l=1$. The model can be 
understood as coming from a parent $3\times 3$ box model. The T-duality 
along 46 of such model is very simple, and produces a set of D3 branes on 
top of a  $\IC^3/(\IZ_3\times \IZ_3)$ singularity, with the generators 
acting as in (\ref{action}). In order to take into account the $\IZ_3$ 
identifications of boxes to transform the parent model into the true 
one, we must quotient $\IZ_3\times \IZ_3$ by the $\IZ_3$ generated by 
$\theta^{-1}\omega$. The three equivalence classes in the quotient are 
$\{1,\theta^2\omega,\theta\omega^2\}$, $\{\theta,\omega,\theta^2\omega^2\}$, 
and $\{\theta^2,\theta\omega,\omega^2\}$, and so the quotient group is 
$\IZ_3$.
 
In order to find its action on $\IC^3$ we have to relate the irreducible 
representations of 
$\IZ_3\times \IZ_3$ with those of $\IZ_3$. A homomorphism sending 
$\RR_{2,1}$ to ${\bf \RR_0}$ is
\beqa
\{\RR_{0,0},\RR_{2,1},\RR_{1,2}\} & \to & {\bf \RR_{0}} \nonumber \\
\{\RR_{1,0},\RR_{0,1},\RR_{2,2}\} & \to & {\bf \RR_{1}} \nonumber \\
\{\RR_{2,0},\RR_{1,1},\RR_{0,2}\} & \to & {\bf \RR_{2}}
\eeqa
(the only other choice, with ${\bf \RR_{1}}$ and ${\bf \RR_{2}}$ 
exchanged, leads 
to identical results). The image of ${\bf 
3}=$$\RR_{1,0}\oplus$$\RR_{0,1}\oplus$$\RR_{-1,-1}$ under this map is 
${\bf 3}=$${\bf \RR_{1}}\oplus$${\bf \RR_{1}}\oplus$${\bf \RR_{1}}$, which 
defines the action of the final $\IZ_3$ on $\IC^3$. This completes the 
construction, showing that the initial brane box model is T-dual to
D3 branes at a $\IC^3/\IZ_3$ singularity.

\medskip

{\bf ii)} Let us consider a more general case as final example. Consider 
the brane 
box model shown in figure \ref{fig:zkshifted}, which consists of a $k 
\times 1$ box model with trivial identification of vertical sides, and 
identifications of horizontal sides accompanied by a shift of $p$ boxes to 
the left. The parent model is given by a $k \times k/\ell$ box model 
(where $\ell$ is the greatest common divisor of $k$ and $p$, 
$\ell=\gcd(k,p)$), whose T-dual is a 
$\IZ_{k}\times \IZ_{k/\ell}$ singularity with generators acting as in 
(\ref{action}) (in this case $k'=1$). The order $k/\ell$ identification is 
taken into account by computing the quotient by the subgroup generated by 
$\theta^{-p} \omega$. There are $k$ equivalence classes, the $i^{th}$ of 
which has the elements $\theta^i (\theta^{-p}\omega)^n$ for $n=0,\ldots, 
k/l-1$. The final group is $\Gamma=\IZ_k$.

Let us find the action on $\IC^3$. A natural homomorphism (fulfilling the 
conditions mentioned above) between the sets 
of irreducible representations is given by $\RR_{i-np,n} \to {\bf 
\RR_{i}}$ for 
$n=0,\ldots,k/\ell-1$, and $0=1,\ldots,k-1$. Under this map the 
representation 
${\bf 3}=$$\RR_{1,0}\oplus$$\RR_{0,1}\oplus$$\RR_{-1,-1}$ becomes ${\bf 
\RR_{1}}\oplus$${\bf \RR_{p}}\oplus$${\bf \RR_{-p-1}}$. This defines the 
action of $\IZ_k$ on $\IC^3$. Notice that the T-duality argument provides 
in a constructive way the type of singularity that we saw in Section~3 
reproduces the starting model.

\bigskip

It is time to revisit an open issue we had in our study of realization of 
field theories using brane box configurations, in Section~2. Namely, 
the fact that different brane configuration can lead to the same field 
theory, the only difference being that e.g. the horizontal fields in one 
appear as vertical or diagonal fields in the other. As we mentioned in 
Section~3 all these brane configurations are reproduced by the same 
singularity simply by changing the correspondence between complex planes 
in the singularity and horizontal, vertical and diagonal fields in the 
brane box model. The T duality argument above improves our understanding of 
the situation. If we start with a singularity 
$\IC^3/\Gamma$, and wish to relate it to a brane box configuration, we 
have to perform T duality along the $U(1)$ orbits in two complex planes, 
say $z_1$, $z_2$. More precisely, by this we mean first substituting the 
singularity by a manifold with the same local behaviour but different 
asymptotics, so that the mentioned orbits have finite radius at infinity, 
and the T dual configuration makes sense \footnote{The situation is 
analogous to the relation of ALE and Taub-NUT metrics.}. The brane 
configuration that arises will be such that the diagonal fields reproduce 
the fields $\Phi_{I,I\oplus A_3}$ (associated to the third complex plane 
$z_3$, {\em i.e.} precisely the non T-dualized one). The fields 
$\Phi_{I,I\oplus A_1}$, $\Phi_{I,I\oplus A_2}$ will map 
to horizontal and vertical fields. These two latter possibilities are 
obviously 
related by the exchange of the roles of NS and NS$'$ branes.

It is then clear that any of the three kinds of fields $\Phi_{I,I\oplus 
A_i}$, $i=1,2,3$, can be taken to reproduce the diagonal fields in a 
T-dual brane box configuration, by merely T-dualizing along the two other 
directions. This means that T dualities of the same singularity along 
different directions reproduce the different brane boxes yielding the same 
four-dimensional field theory. This is a nice result, since it points towards 
some unifying description of all the brane box models yielding the same field 
theory.

\subsection{T-duality of wrapped NS fivebranes}

In our previous arguments showing the T-duality relation between the brane 
box configurations and the D3 branes at singularities, the role played by 
the D branes is quite trivial. We can consider removing them from the 
picture, and look at the result we have obtained as a T-duality between 
certain grids of 
intersecting NS fivebranes  wrapping cycles in a torus, and certain 
non-compact Calabi-Yau threefold geometries. The latter can be roughly 
described as singularities of the type $\IC^3/\Gamma$ with modified 
asymptotics that make two of the coordinates ($x^4$ and $x^6$) compact at 
infinity.

Such grids have been described as infinite grids on the plane modded out 
by certain translations, giving rise to identifications of the sides of 
some unit cell. When the identifications are trivial, the NS fivebranes in 
the grid wrap cycles of type $(1,0)$ and $(0,1)$ in the torus. When the 
identifications are non-trivial, the NS fivebranes wrap more complicated 
cycles in the torus. It is interesting to translate the specific infinite 
grids we have been studying to the cycles the fivebranes wrap when one 
effectively restricts to the quotient torus.

\begin{figure}
\centering
\epsfxsize=5in
\hspace*{0in}\vspace*{.2in}
\epsffile{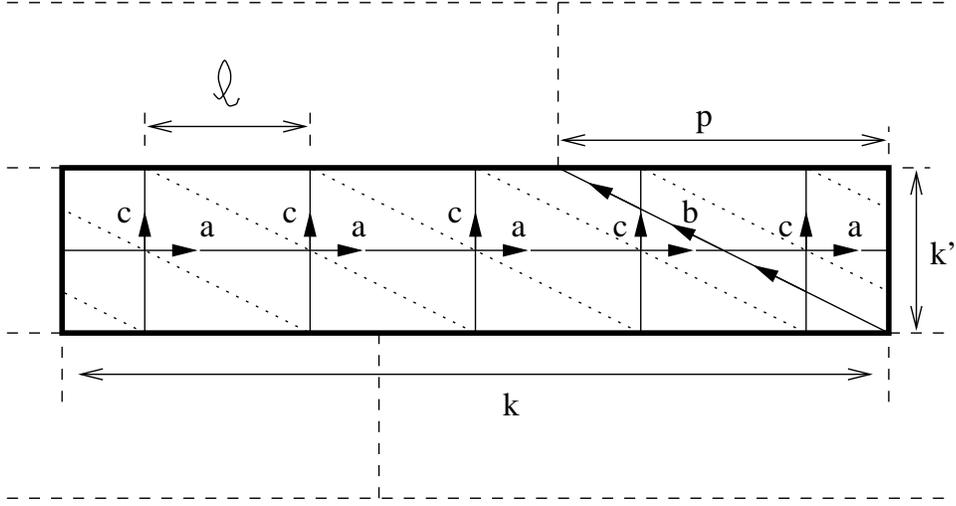}
\caption{\small The unit cell in a model with $k=5\ell$, $p=2\ell$. The 
cycle of 
the torus wrapped by the NS$'$ branes is the horizontal line labeled 
`$a$'. The cycle `$c$' corresponding to the NS branes wraps the 
vertical direction several times due to the shifted identifications (suggested 
by the dotted lines). The cycle `$b$', corresponding to the slanted line, 
is the dual to `$a$'.}
\label{fig:wrap1}
\end{figure}

One can always define the cycle wrapped by a particular kind of brane, say 
the NS$'$, to be of type $(1,0)$. This amounts, in the language of 
infinite grids, to saying that one can always choose a unit cell with 
trivial identifications of, say, vertical sides. So let us consider the 
most general such configuration, by now familiar, consisting on a 
$k\times k'$ box model with trivial identifications of vertical sides and 
identification of horizontal sides accompanied by a shift of $p$ boxes to 
the left. The picture corresponding to the following explanations is 
depicted in figure \ref{fig:wrap1} for a particular example.

The NS$'$ branes wrap a $(1,0)$ cycle which 
we denote by $a$. The dual cycle, of type $(0,1)$ is denoted $b$, and is 
represented by a slanted line, closed due to the shifted identification. 
The NS brane wraps a cycle $c$, represented as a set of vertical lines 
which form a closed loop due to the shifted identifications (suggested by 
dotted lines). This cycle can be expressed in terms of the basic homology 
cycles, $c=na+mb$. We can
 determine the type $(n,m)$ of the cycle $c$ that the NS branes are 
wrapping by simply 
looking at its intersection number with the basic cycles $a$ and $b$, 
$c\cdot a=-m$, $c\cdot b=n$. Recall that the intersection number of two 
cycles, $c_1\cdot c_2$, counts the number of their intersection 
points, with `plus' signs when the orientation defined by $c_1$, $c_2$ 
(in this order) is positive, and `minus' signs otherwise. Notice that 
$a\cdot b=1$.
Due to the shifted identification, a {\em single} NS brane corresponds to 
$k/\ell$ vertical lines in the unit cell, where $\ell=\gcd(k,p)$. Thus we 
have $c\cdot a=-k/\ell$. Noticing that 
the vertical lines in the unit cell have an equal spacing of $\ell$ 
boxes, we also have $c\cdot b=p/\ell$.
Thus, the NS branes wrap cycles of type $(p/\ell,k/\ell)$.

Other choices of the unit cell, for example one with trivial 
identification of {\em vertical} sides, yield other labelings of the 
same cycles, but they are simply related by a $SL(2,\IZ)$ transformation 
on the complex structure of the torus.

\begin{figure}
\centering
\epsfxsize=3.5in
\hspace*{0in}\vspace*{.2in}
\epsffile{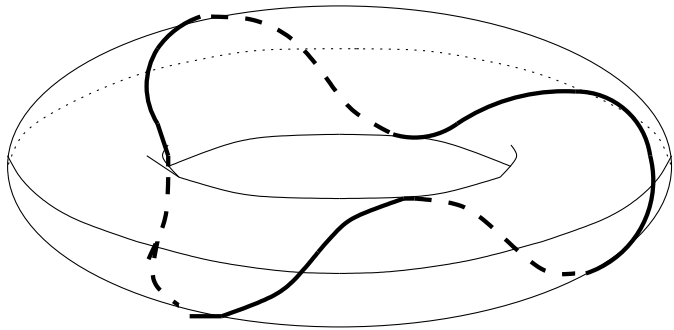}
\caption{\small A pictorial representation of a cycle of type $(1,3)$, 
which is 
wrapped by the NS branes in the $3\times 1$ box model of figure 
\ref{fig:z3}.}
\label{fig:wrapped}
\end{figure}

As a simple example, we can consider the $3\times 1$ box model depicted in 
figure \ref{fig:z3}. The NS$'$ brane wraps a $(1,0)$ cycle, and the NS 
brane wraps a $(1,3)$ cycle. A pictorial version of this model is shown in 
figure \ref{fig:wrapped}.

So our considerations in the preceding sections show how to perform 
T-duality in the following type of configuration: two sets of 
fivebranes, one spanning 01235 and an arbitrary cycle in a two-torus, 
another spanning 01237 and another arbitrary cycle in the torus. After 
T-duality in the two directions of the torus, one obtains a certain 
manifold 
which for many purposes can be approximated by a $\IC^3/\Gamma$ 
singularity. The orbifold group $\Gamma$ is determined from the grid of 
fivebranes by the 
recipe presented in sections 4.1, 4.2, {\em i.e.} from considerations 
concerning the four-dimensional theory on D-brane probes of the 
configuration.
This is an amusing aspect of the present study.
By looking at two different configurations which lead to four dimensional
supersymmetric gauge theories, one is led to these T-dual pairs.
The methods presented here actually demonstrate how,
starting with one configuration, we can use four dimensional
gauge theories to calculate the T-dual partner.

\section{Gauge couplings and AdS/CFT correspondence}

\subsection{The marginal couplings}

We have already mentioned in Section 2 that the four-dimensional $\NN=1$ 
gauge theories we are considering have a certain number of 
marginal couplings; there is a manifold of renormalization 
group fixed points in the space of couplings. In 
\cite{hsu} it was shown that one such marginal parameter existed for each 
independent horizontal row of boxes in the brane box configuration, 
another for each independent vertical column of boxes, and another for 
each independent line of boxes running diagonally from upper right to 
lower left.
Recall that the `vertical' parameters were interpreted in the brane box 
model as the independent distances between NS branes. Similarly, the 
`horizontal' couplings corresponded to the
independent distances between NS$'$ branes. The interpretation of the 
`diagonal' parameters is less clear, even though they seem to be related to 
fields living at the intersection of NS and NS$'$ branes. The overall coupling 
is determined by the area of the torus parametrized by 4,6.

We would like to achieve some understanding of these field theory 
parameters in the construction {\em via} branes at singularities.
A quite general family of models where we can study this issue is the 
field theories obtained as $k\times k'$ box models with trivial 
identifications of sides. These theories have one overall coupling, $k-1$ 
`vertical' marginal couplings, $k'-1$ `horizontal' couplings, and $r-1$ 
`diagonal' ones, where $r=\gcd(k,k')$ as usual.

The singularity that 
reproduces this field theory in the world-volume of D3 brane probes is 
$\IC^3/\IZ_{k}\times \IZ_{k'}$, with the generators $\theta$, $\omega$ 
acting on $\IC^3$ as in (\ref{actionzkzkp}). 

We would like to identify how 
these parameters are encoded in the singularity. The overall 
coupling is given by the string coupling in the usual way. In order to 
understand the remaining paremeters, 
it is useful at this point to recall the case of the $\NN=2$ $SU(N)^k$ 
theories. These models have $k$ marginal couplings, interpreted as one 
overall and $k-1$ `vertical' couplings in the brane box construction. 
These same parameters are interpreted in the T-dual picture of D3 branes 
at an $A_{k-1}$ singularity as the string coupling, and the $k-1$ 
integrals $a_i, v_i$ of the NS-NS and RR two-forms over the two-cycles 
of the resolved ALE space, equations (\ref{flux1}), (\ref{flux2}).
Since our $\NN=1$ gauge theories have flat directions connecting them to 
$\NN=2$ theories (by removing either kind of NS fivebranes in the brane 
box picture), the parameters are expected to be also encoded as integrals 
of two-forms over two-cycles implicit in the singularity. So we should 
understand some basic features about the resolution of $\IC^3/(\IZ_k\times 
\IZ_{k'})$ singularities. 

Singular points in the quotient appear from 
points in $\IC^3$ which are left invariant under some element of the 
discrete group. We can distinguish several types of them. First, there is 
the (complex) curve defined by $z_1=z_3=0$, and parametrized by $z_2$, 
which is invariant under the $\IZ_k$ subgroup generated by $\theta$. In 
the quotient it becomes
a curve of $A_{k-1}$ ALE singularities. This set of singularities is 
precisely the only one remaining when in the T-dual brane box model the 
NS$'$ branes are removed, and one has Higgssed the theory to an $\NN=2$ 
$SU(N)^k$
model. So, it is natural to associate the $k-1$ distances between NS 
branes (the only parameters that remain in this Higgs branch) to the 
integrals of the two-form fields over the $k-1$ independent two-cycles 
that 
resolve the singularity. Analogously, there is another curve defined by 
$z_2=z_3=0$, which is left fixed by the $\IZ_{k'}$ subgroup generated by 
$\omega$, and which becomes a curve of $A_{k'-1}$ singularities in the 
quotient. The integrals of the two-form fields over the corresponding 
two-cycles encode the distances between the NS$'$ branes, since these are 
the only parameters remaining on the Higgs branch associated to the 
removal of the NS branes in the brane box picture, which yields a 
$SU(N)^{k'}$ $\NN=2$ field theory.

But there is more to the story. There is yet another curve of 
singularities. It corresponds to $z_1=z_2=0$, which is left invariant by a 
$\IZ_r$ subgroup (with $r=\gcd(k,k')$, as before), 
generated by $\theta^{k/r} \omega^{-k'/r}$, whose action on $\IC^3$ is
\beq
(z_1,z_2,z_3) \to (e^{2\pi i/r}z_1,e^{-2\pi i/r}z_2,z_3)
\eeq
This becomes a curve of $A_{r-1}$ singularities in the quotient. It is 
easy to see that the field theory has a flat direction connecting it to a 
$SU(N)^r$ $\NN=2$ gauge theory. This breaking, however, is not manifest in 
the brane box construction \footnote{However, as in section 2, one can 
consider another brane box configuration yielding the same field theory, 
and in which this Higgs branch is manifest.}, and that is the main reason 
the corresponding 
$r-1$ parameters were not fully understood in the brane box configuration. 
In the singularity language,
however, the integral of the two-forms over the two-cycles resolving the 
$A_{r-1}$ singularity are the natural candidates for the remaining $r-1$ 
parameters. The symmetry between the three kinds of breaking to $\NN=2$ 
in the field theory is manifest in the singularity picture as the 
symmetry between the three complex planes in $\IC^3$. This very nice 
result provides a geometrical understanding of all the parameters in the 
gauge theory, and may help in their interpretation in the brane box 
language.

We should be aware that the resolution of the singularity has not been 
completed yet. The origin in $\IC^3$ is left invariant by all the elements 
in $\Gamma$, and the corresponding singularity in the quotient requires 
further blow-ups. Consequently, the integrals of p-form fields over the 
resulting cycles seem to increase the number of parameters in 
the model. However, there is no contradiction with the above statement 
that the model contains $k+k'+r-2$ {\em independent} couplings. The 
complete 
space of couplings is certainly larger, but in order to have a 
conformal theory, so that microscopic couplings exist, the couplings must 
lie in a ($k+k'+r-2$)-dimensional 
manifold. It is quite a remarkable fact that these independent parameters 
are precisely the integrals $a_i$, $v_i$ of the two-forms over the 
non-compact divisors (those resolving the curves of singularities, rather 
than the singularity at the origin). The integrals over the remaining 
cycles are (possibly complicated) functions of these, and do not provide 
{\em new} independent couplings.

In the following section we use an argument based on the recent conjecture 
relating 
large $N$ gauge theories to string theory on Anti de Sitter spaces to 
support our identification of the marginal parameters.

Finally, we would like to stress that even though the agreement in the 
counting of marginal couplings has been shown
only for a certain class of models, namely when $\Gamma=\IZ_k\times 
\IZ_{k'}$, the argument also works for other Abelian quotient 
singularities. Actually, there is a direct relation between closed lines 
of boxes in the brane box diagram, and subgroups of $\Gamma$ which leave 
invariant a complex curve in $\IC^3$. It would also be nice to 
extend these results to more general subgroups of $SU(3)$.

\subsection{The AdS/CFT correspondence}

In this subsection we connect the discussion of the previous one with the 
recent conjecture relating the large $N$ limit of gauge theory to string 
theory on a certain background \cite{malda,witads,gk,kl,gkp}. The aim of 
the argument is to find out the number of marginal operators of the 
four-dimensional $N=1$ gauge theory. The theories on the world-volume of 
D3 branes located at orbifold 
singularities $\IC^3/\Gamma$ were proposed in \cite{ks,lnv} (see also 
\cite{bkv,bj,k2,k1}) as simple models to study 
gauge theories with reduced or with no supersymmetry which were conformal, 
at least in the large $N$ limit, by using the connection with 
supergravity/string theory on the space $AdS_5\times S^5/\Gamma$. The 
basic requirement for such $\Gamma$ is that it should act only on $S^5$, 
so that the nice property that the group of isometries of the $AdS$ space 
becomes the conformal symmetry on the boundary (where, roughly, the gauge 
theory lives) is preserved. Within this class of theories, the detailed 
correspondence between fields propagating on the $AdS_5$ and operators on 
the boundary, analyzed in \cite{witads,gkp}, carries over and can be 
applied directly. As in the maximally supersymmetric case, the relation 
between the mass $m$ of $p$-form field in $AdS_5$ and the conformal 
dimension $\Delta$ of the operator it couples to in the boundary is
\beq
(\Delta+p)(\Delta+p-4)\,=\,m^2
\label{massdim}
\eeq
One can then hope to be able to compute the conformal dimensions of 
primary chiral operators in the conformal theory by computing the 
Kaluza-Klein reduction of  ten-dimensional Type IIB supergravity on 
$S^5/\Gamma$ to find the masses of fields propagating on $AdS_5$, in 
parallel with the comparison made in \cite{witads} for the $\NN=4$ case. 
In \cite{ozter} this computation was partially performed by taking 
the KK excitations on $S^5$ and performing a projection onto 
$\Gamma$-invariant states.

However we must stress that this procedure may not give the complete 
answer, since it only takes into account the untwisted modes in the 
quotient. Any possible twisted mode is completely missed by the 
supergravity approximation, and will be manifest only when the full string 
theory on the orbifold is considered. This is a very interesting issue, 
since it will provide indications of how the stringy modes enter the 
conjectures relating gauge theory and string theory.

Twisted modes appear when $S^5/\Gamma$ contains singularities, the 
structure of which is found by looking for fixed points of the action of 
$\Gamma$ on $S^5$. To this end it will be useful to realize the $S^5$ as 
the unit five-sphere in an auxiliary $\IC^3$ parametrized 
by $(z_1,z_2,z_3)$ 
\beq
|z_1|^2+|z_2|^2+|z_3|^2\,=\,1
\eeq

The main observation, already made in \cite{ks}, is that the elements of 
$\Gamma$ whose only fixed point is the origin of this $\IC^3$, act 
freely on the $S^5$, and do not induce singularities in the quotient. The 
elements in $\Gamma$ that leave fixed a complex curve in $\IC^3$, however, 
will induce singularities on the quotient $S^5/\Gamma$. In our 
$\Gamma=\IZ_k\times \IZ_{k'}$ example the action of $\Gamma$ on 
this $\IC^3$ is as in (\ref{actionzkzkp}). The curve $z_1=z_3=0$ of fixed 
points in $\IC^3$ intersects the unit five-sphere along the $S^1$ given by 
$|z_2|^2=1$. This induces a real curve of $A_{k-1}$ singularities in the 
quotient $S^5/\Gamma$. Analogously, there is another $S^1$, given by 
$|z_1|^2=1$, of $A_{k'-1}$ singularities, and another $S^1$, $|z_3|^2=1$, 
of $A_{r-1}$ singularities. These real curves are disjoint on the $S^5$, 
so there are no further singularities.

Even though the supergravity description is not valid, these 
singularities are harmless in the full string theory, and there 
are some states appearing as twisted sectors. The massless twisted fields 
at each of these $\IZ_n$ orbifold singularities will be those appearing in 
Type IIB compactification on $A_{n-1}$ ALE spaces. Namely, 
there will be $(n-1)$ sets of fields, each containing a two-form 
and five scalars. The self-dual two-forms appear from 
the integral of the Type IIB four-form over each of the $n-1$ two-cycles in 
the resolution of the singularity, two of the scalars from the integrals 
of the RR and NS-NS two-forms over the two-cycles, and the remaining three 
scalars from the positions of the corresponding centers in the ALE metric. 
 From the three kinds of singularities, we get $(k-1)+(k'-1)+(r-1)$ sets of 
such fields.
These fields are massless and propagate in $AdS_5\times S^1$, where $S^1$ 
is the corresponding circle of singularities. So one obtains a tower of 
states propagating on $AdS_5$, associated to the Kaluza-Klein reduction of 
these six-dimensional fields on the $S^1$. It would be interesting to 
match the 
masses of these modes with the conformal dimensions of certain operators 
on the boundary theory. We will do so for the massless scalar modes in 
$AdS_5$ operators, leaving the general question for future research. 

Let us first discuss the complex scalars coming from the integrals of the 
$B$-fields over the collapsed two-cycles. These are massless scalar 
fields, so from (\ref{massdim}) we see they must couple to 
marginal operators in the conformal field theory. We had already counted 
and identified them. They are given by $\sum_{i} tr F_i^2$, where 
the sum runs over the group factors associated to boxes forming 
independent horizontal, vertical and diagonal lines in the corresponding 
brane box diagram. Their number is thus 
$(k-1)+(k'-1)+(r-1)$, precisely the number of massless scalars of the type 
mentioned. From our analysis of branes at singularities in the previous 
section, we also infer that the appropriate couplings between the bulk 
fields and the boundary operators exist, {\em i.e.} the fields play the 
role of coupling constants for the gauge theory.

It is thus a fortunate circumstance that there are no further 
singularities on $S^5/\Gamma$. Otherwise the integrals of $p$-forms over the 
new cycles would have provided further 
massless scalar fields propagating on $AdS_5$. This would require the 
theory to have more marginal couplings, a fact which is not found in the 
field theory analysis. The argument above thus provides supporting 
evidence for our counting and identification of the independent parameters 
in the gauge theory. Even though the $AdS$ argument is only valid for 
large $N$, our identification of the parameters with the integrals of 
B-fields in section~5.1 was mainly based on field theory properties valid 
for all $N$ (namely, Higgs branches breaking to $\NN=2$).

As for the three remaining scalar modes, we see that in the $\NN=2$ case 
they transform as a triplet of $SU(2)_R$.
So they couple to the D-terms of the gauge theory. For the $\NN=1$ 
theories, these modes couple to whatever operators become the D-terms 
after the appropriate breaking to $\NN=2$.

\medskip

We finish this section with some side comments our study of orbifold 
theories suggests.
 
The analysis of marginal couplings in $\NN=2$ theories is simple, and 
can be extended to non-abelian discrete groups $\Gamma$ as well.
In all the quotients $S^5/\Gamma$, with $\Gamma$ and ADE subgroup of 
$SU(2)$, there will be singularities and twisted sectors.
The number of two-cycles in the resolution of the 
singularity is given by the number of nodes in the corresponding 
Dynkin diagram. This is also the number of factors in the gauge 
group, and thus also the number of marginal coupling of the theory (not 
counting the overall coupling). So again the twisted sector modes are the 
appropriate fields in $AdS$ to account for certain operators in the gauge 
theory. 

Once these techniques have shown the geometrical features in the 
$AdS$ picture that underlie the existence of marginal couplings in the gauge 
theory, we can use such knowledge and apply it even to non-supersymmetric 
models in the large $N$ limit. It is known \cite{ks,lnv} that 
non-supersymmetric theories obtained from D3 branes on top of a 
$\IR^6/\Gamma$ singularity (with $\Gamma$ a generic subgroup of $SU(4)$) 
have at least one marginal coupling, which corresponds to the massless 
dilaton in the $AdS$. Now we see that if $\Gamma$ has (real) curves of fixed 
points on the $S^5$, yielding ALE singularities in the quotient, the 
non-supersymmetric theories will have new marginal 
operators in the large $N$ limit. As an example consider a $\IZ_{10}$ 
singularity, with generator $\theta$ acting on the R-symmetry 
quantum numbers of the fermions (in the ${\bf 4}$ of $SU(4)_R$) through 
the representation
\beqa
{\bf 4}=\RR_{1}\oplus \RR_{1}\oplus\RR_{2}\oplus\RR_{-4}.
\eeqa
The action on the R-symmetry representation ${\bf 6}$ of the bosons is 
\beqa
{\bf 6}=\RR_{2}\oplus\RR_{-2}\oplus\RR_{3}\oplus\RR_{-3}\oplus
\RR_{3}\oplus\RR_{-3}
\eeqa
The only singularity in $S^5/\Gamma$ comes from the 
fixed points of $\theta^5$, and that it is of $A_1$ type. Thus we expect 
this non-supersymmetric theory to have {\em two} marginal couplings. It 
should not be difficult to construct further examples along this line.

\subsection{Strong coupling limits in the gauge theory}

One of the interesting points about the identification of parameters we 
have carried out is that it allows for the comparison of some dynamical 
field theory phenomena in the brane box and the singularity pictures. As 
an example, we briefly comment on the limit in which some of the 
independent parameters in the theory go to zero \footnote{We are 
thankful to M.~J.~Strassler for discussions on the following arguments.}.
It is important to note that the following discussion is valid only to
finite gauge theories. For such models, the branes are not bent and the
position of the NS branes are good parameters.

Limits with vanishing parameters are obtained in the brane box picture by
letting several, say 
$n$, NS branes coalesce. This corresponds to setting to zero $n$ of the 
`vertical' parameters, and is associated to a strong coupling limit for 
some of the gauge factors. The most relevant feature of this limit is the 
appearance of a six-dimensional $U(n)$ gauge symmetry in the world-volume
of the NS branes \footnote{To be precise, in order to get this enhanced 
symmetry one should also tune the Wilson lines of the world-volume gauge 
fields along $x^4$. Thus the enhanced symmetry locus is reached upon 
tuning $n$ {\em complex} parameters. There are additional parameters
corresponding to 89 positions of the NS branes but they are set to zero in
a typical construction.}. This is interpreted as an enhanced 
global symmetry from the point of view of the four-dimensional gauge 
theory.

One can recover this behaviour in the singularity picture by explicit 
mapping (via T-duality) of the parameters involved. We have mentioned that 
the distance between NS branes (and the corresponding Wilson line degrees 
of freedom) are mapped to the integrals of the Type IIB two-forms over 
two-cycles implicit in the $A_{k-1}$ singularity. The strong coupling 
limit we have discussed corresponds to setting these B-fields to zero. In 
this regime, D3 branes wrapping the two-cycles give rise to tensionless 
strings. Notice that one of the six dimensions in which this theory lives, 
$x^6$ is compact, and T-dualizing along it we recover the picture of gauge 
symmetry enhancement we had in the brane box construction.

The picture of the strong coupling limit in the singularity language can 
be translated to the AdS picture without much change, using the 
information we obtained in 
section~5.2. In such a strong coupling limit, tensionless strings appear
propagating on $AdS_5\times S^1$. The modes propagating on $AdS_5$ are 
obtained by mode expansion on the `internal' $S^1$. The massless modes in 
$AdS_5$ are a multiplet of $U(n)$ gauge bosons, which arise from the 
tensionless string wrapping the $S^1$. The gauge symmetry in the bulk is 
interpreted as a global symmetry on the boundary field theory; the 
massless fields couple to the corresponding conserved currents on the 
boundary.

This example illustrates how the T-duality we have established may help 
in understanding other constructions. Without the intuition provided by 
the brane box configurations, the enhanced global symmetry observed from 
the $AdS_5$ argument would have been harder to interpret. On the other 
hand, the singularity picture may help in understanding some interesting 
regimes not so intuitive in the brane box picture. For example, those 
related to setting to zero some diagonal parameters.

\section{Non-conformal theories}

In this section we explore the singularity picture corresponding 
to brane box models with different number of D5 branes in each 
box. The basic ingredients -- fractional branes -- that enter the 
definition of the corresponding
configurations of branes at singularities have appeared mainly in the 
context of D0 branes and M(atrix) theory \cite{douglas,distler,diaconescu}, 
without any reference to configurations of intersecting branes. We will 
argue these type of objects provide the T-dual of the brane box 
configurations with non-constant number of D5 branes. Such relation was 
explored in \cite{lust} for the case of $\NN=2$ theories.
Other related issues in models with $N=1$ supersymmetry were discussed in
\cite{brodie}. 

\subsection{Fractional branes}

The first relevant observation is that the T duality  relation 
between the grid of fivebranes and the singularity does not depend on the 
distribution of D branes, so the recipe of sections~3 and 4, that relates 
a given grid to a given singularity (and vise versa), remains valid.
Thus, starting with a given brane box configuration we can determine the 
orbifold group $\Gamma$ of the singularity picture. We also know how to 
associate each box with an irreducible representation of $\Gamma$. In the 
following it will be convenient to label the boxes by their corresponding 
irreducible representation.

The information about the number $n_I$ of D5 branes in the box labeled 
$\RR_I$ is encoded in the singularity picture in how the orbifold group 
acts on the Chan-Paton indices of 
the T-dual D3 branes. If $n_{tot}$ denotes the total number of D5 branes 
in the brane box configuration, $n_{tot}=\sum_{I} n_I$, the T-dual 
configuration can be described as an orbifold of $\IC^3$ with $n_{tot}$ D3 
branes in the covering space. Here the counting includes all the copies 
under the orbifold action, if present. The action of 
$\Gamma$ on the Chan-Paton factors is defined by a $n_{tot}$-dimensional 
representation. The adequate choice to reproduce the spectrum in 
the brane box configuration is
\beqa
\RR_{C.P.}=\bigoplus_{I} n_I \RR_I
\eeqa
as we will show below.
Observe that when the number of D5 branes on each box is the same, say 
$N$, this representation consists of $N$ copies of the regular 
representation $\RR_{\Gamma}\equiv \bigoplus_I \RR_I$, as should be the 
case.

The spectrum is determined following the rules in \cite{lnv}. It is easy 
to see that it reproduces the spectrum of the field theory obtained 
in the brane box picture. The gauge group is $\prod_{I} SU(n_I)$. 
There are three kinds of chiral multiplets for each $I$, whose 
gauge quantum numbers are determined by computing the tensor products 
${\bf 3}\otimes \RR_I$. There are fields, which we denote by 
$\Phi_{I,I\oplus A_1}$, 
transforming in the $(\fund,\antifund)$ of $SU(n_I)\times 
SU(n_{I\oplus A_1})$. 
Similarly, the fields $\Phi_{I,I\oplus A_2}$ transform in the 
$(\fund,\antifund)$ of $SU(n_I)\times SU(n_{I\oplus A_2})$, and 
$\Phi_{I,I\oplus A_3}$ transform in the $(\fund,\antifund)$ of $SU(n_I)\times 
SU(n_{I\oplus A_3})$. Here it is understood that if some $n_I$ vanishes 
the 
corresponding group, and the chiral multiplets charged under it, are not 
present.

The basic building block of these configurations are, in the brane box 
picture, models with one D5 branes in one box (say, labeled $\RR_I$) 
and zero in the rest \footnote{These configurations violate the restrictions 
on the 
numbers of D5 branes derived in \cite{gg}. Since for the moment we are 
treating these configurations merely as building blocks, we will ignore 
this difficulty.}. Correspondingly, there are some basic configurations 
in the singularity picture, which correspond to a choice of Chan-Paton 
factors in the representation $\RR_{C.P.}=\RR_I$ (notice that $n_{tot}=1$ in 
these configurations). The D-brane described by these Chan-Paton 
factors is 
called a `fractional brane'. There are different kinds of these objects, 
each one being characterized by the representation $\RR_I$ of its 
Chan-Paton factors. Their name is due to 
the observation that a combination of such branes, one for each 
irreducible representation of $\Gamma$, has Chan-Paton factors 
$\RR_{C.P.}=\bigoplus_{I}\RR_I\equiv \RR_{\Gamma}$ and has the 
interpretation of a (whole) D3-brane in the quotient.

 From the rules above, one can determine the world-volume field theory of 
such configuration \footnote{For simplicity we will discuss in the
classical limit, where even a single such brane is dynamical, its 
world-volume $U(1)$ gauge group not being frozen. The discussion extends 
straightforwardly to other configurations.}. It has no flat directions, 
and so the branes are 
stuck at the singular point. This can also be understood by noticing that 
in the flat cover $\IC^3$ of the orbifold we have only one D3 brane, and 
the only $\Gamma$-invariant configurations corresponds to placing it at 
the origin. 
This last argument makes it clear that models with several 
fractional branes may allow for $\Gamma$-invariant configurations with 
branes away from the origin. In the quotient, the corresponding brane 
will be able to move away from the singularity. The clearest example is 
having one fractional brane of each kind, $\RR_{C.P.}=\bigoplus_I 
\RR_I\equiv\RR_{\Gamma}$, which defines a brane that can move freely in 
the quotient space $\IC^3/\Gamma$. The world-volume field theory contains 
the appropriate Higgs branches. Actually, these are clearly visible in the 
brane box construction. The configuration has one D5 brane in each box, so 
that they can recombine and leave the grid of NS and NS' along $x^5$, 
$x^7$, $x^8$, $x^9$ (additional moduli are provided by the Wilson lines 
around 4 and 6 of the worldvolume gauge fields).
These are the types of objects we have been considering in previous subsections.

In some cases, which will be our main interest in forthcoming 
considerations, there may be certain combinations of the basic fractional 
branes which are allowed to move away on a submanifold of $\IC^3/\Gamma$. 
This type of motion will occur when, in the brane box configurations, we 
have the same number of D5 branes in each box belonging to {\em e.g.} a 
given horizontal row. In such case, the D5 branes in the row can recombine 
and move away along $x^7$, stretched between NS' branes. To make the 
discussion of the singularity picture clearer, we can consider a $k\times 
k'$ box model with trivial identifications, even though the conclusions hold 
in other cases as well. The configuration with one D5 brane in the boxes 
belonging to the $j^{th}$ row is mapped to a set of fractional branes 
defined by $\RR_{C.P.}=\bigoplus_i \RR_{i,j}$. The flat direction in the 
worldvolume field theory implies that this set of fractional branes is 
allowed to move 
along the curve of $A_{k'-1}$ singularities in the quotient, but not away 
from it.

There exists an analogous set of fractional branes defined by 
$\RR_{C.P.}=\bigoplus_j \RR_{i,j}$, which is T-dual to a configuration 
with one D5 brane in the boxes belonging to the $i^{th}$ column, and zero 
in the others. There is a flat direction in the field theory which allows 
the D5 branes in the box model to recombine and move away along $x^5$. 
This is mapped to moving the set of fractional D3 branes along the curve of 
$A_{k-1}$ singularities. 

Finally, there is a set of fractional branes 
given by $\RR_{C.P.}=\bigoplus_{l=1} \RR_{i+l,j+l}$, which is T-dual to a 
brane box 
configuration with one D5 brane in all boxes on the diagonal of the 
box $(i,j)$. Even though it is not obvious in the brane box construction, 
the field theory contains a flat direction, which corresponds to moving 
the set of D3 branes along the $A_{r-1}$ curve.

\medskip

A geometric interpretation of the fractional branes has been proposed in 
\cite{polchinski,douglas,dgm}, as higher dimensional branes (or bound 
states thereof) which are 
wrapping the cycles which are implicit in the singularity of the orbifold.
For example, in the case of $A_{n-1}$ ALE singularities, the $(n-1)$ basic 
kinds 
of fractional branes (labeled by $\RR_i$, for $i=1,\ldots,n-1)$) can be 
understood as some sort of D5 branes wrapping 
the $(n-1)$ independent two-cycles $\Sigma_{i,i+1}$ which resolve the 
singularity. The fractional brane corresponding to $\RR_{0}$ is associated 
to the cycle represented by the affine node in the extended Dynkin 
diagram (this cycle is, homologically, minus the sum of all the rest). The 
homology relation between the $n$ cycles explains the fact that a set 
of $n$ fractional branes represents a whole D3 brane in the quotient, 
which wraps no cycle.

The main reason for this interpretation is the fact that a fractional 
brane couples to the closed string modulus which controls the blow-up 
parameters of the corresponding two-cycle. This analysis has been 
partially extended to the case of $\IC^3/\Gamma$ singularities 
\cite{dgm}, where the fractional branes are understood as D5 and D7 
branes 
wrapping the two- and four-cycles implicit in the singularity. However, 
as far as we know there is no systematic way of associating a given 
irreducible representation with a given cycle. It would be interesting 
to develop such geometrical interpretation, but we will not pursue this 
issue in the present work. 
Rather, in the following subsection we will center on a (quite large) 
family of models for which such geometric interpretation is simple. 

\subsection{`Sewing' $\NN=2$ models}

\begin{figure}
\centering
\epsfxsize=6in
\hspace*{0in}\vspace*{.2in}
\epsffile{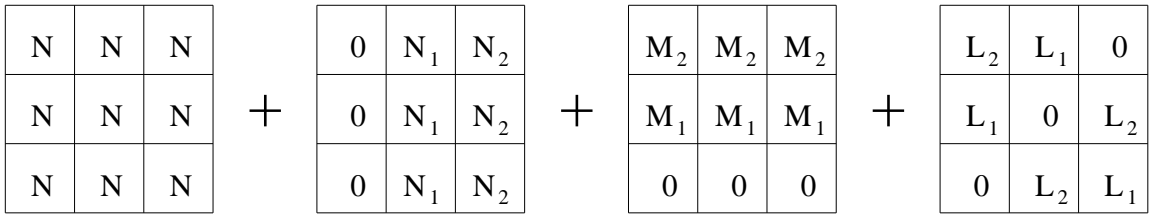}
\caption{\small `Sewing' $\NN=2$ models: A large family of theories can be 
obtained by adding together models formed by whole rows, columns and 
diagonal lines of boxes. Here the numbers denote the number 
of D5 branes in the box.}
\label{fig:threestep}
\end{figure}

The construction of the models we are to consider is as follows. We start 
with any desired grid of NS and NS' branes, with equal number $N$ of D5 
branes. For concreteness we will speak in terms of a $k\times k'$ box 
model with trivial identifications, but the construction is possible in 
the general case. In the singularity picture, we have $\Gamma=\IZ_k\times 
\IZ_{k'}$, and $\RR_{C.P.}=N\bigoplus_{i,j} \RR_{i,j}$
The construction proceeds in three steps, which are depicted in figure 
\ref{fig:threestep} for a $3\times 3$ case (with trivial identifications).
 
The first step is to add $N_i$ branes to each of the boxes belonging to 
the $i^{th}$ column in the grid ($i=1,\ldots, k-1$)\footnote{Notice that 
it is redundant to allow for $N_0$ D5 branes along the $0^{th}$ column, 
since one could reabsorb this in a redefinition of $N$ and the $N_i$'s. 
A similar comment applies in the following steps of the construction.}. 
$N_i$ is kept constant 
within a column, but varies from one column to another. In the singularity 
picture, we have added some fractional D3 branes, which are described by 
$\RR_{C.P.}=\bigoplus_{i,j} N_i \RR_{i,j}$. The worldvolume field theory 
has flat directions which, in the brane box picture, correspond to moving 
whole columns of D5 branes 
along $x^5$, or in the singularity picture, to moving sets of fractional 
branes along the curve 
of $A_{k-1}$ singularities. The configuration of D branes in the 
singularity is geometrically interpreted as having $N_i$ D5 branes 
wrapping the $i^{th}$ two-cycle in the resolution of the $A_{k-1}$ 
singularity, and $N$ D3 branes free to move in the bulk.

The second step is adding $M_j$ D5 branes to the boxes belonging to the 
$j^{th}$ row. In the singularity picture, the new fractional branes we 
have added have Chan-Paton factors $\RR_{C.P.}=\bigoplus_{i,j} M_j 
\RR_{i,j}$. The geometric interpretation of this set in the singularity 
picture is having $M_j$ D5 branes wrapping the $j^{th}$ two-cycle in the 
resolution of the $A_{k-1}$ singularity. These add to the brane we 
had before. There are two kinds of flat directions, moving either whole 
rows along $x^7$, or whole columns along $x^5$. They are mapped to the 
independent motions of each kind of fractional brane along the curves of 
$A_{k'-1}$ and $A_{k-1}$ singularities.

An interesting feature of the models we have obtained after these two 
steps is that they provide the most 
general solution to the constraints derived in \cite{gg}. These were 
obtained by considerations on the bending of the NS fivebranes in the 
brane box model. They state that the numbers $n_{i,j}$ of D5 branes at 
the box in the position $i,j$ have to fulfill the ``sum of 
diagonals rule'', equation (\ref{sodr}),
\beq
n_{i,j} + n_{i+1,j+i} = n_{i,j+1} + n_{i+1,j}
\label{sodr2}
\eeq
for all $i,j$. The most general solution to these conditions can be 
written as
\beq
n_{i,j}=n_{i,0}+n_{0,j}-n_{0,0}
\label{gralsol}
\eeq
Equation (\ref{sodr2}) is a simple difference equation and the solution
(\ref{gralsol}) is obtained by a double summation over the indices $i$ and $j$.
Defining $N_i=n_{i,0}-n_{0,0}$, $M_j=n_{0,j}-n_{0,0}$, and $N=n_{0,0}$, we 
can recast (\ref{gralsol}) as
\beq
n_{i,j}=N+N_i+M_j
\eeq 
This is precisely the structure of our models, where the number of D5 
branes in a box is controlled by the row and column it belongs to.

The claim in \cite{gg} is that these are the most general gauge theories
that can be realized in the brane box setup. However, notice that in the 
singularity picture there is a further curve of singularities, around 
whose two-cycles we can wrap some fractional branes. This is a possibility 
suggested by the symmetry of the three curves of singularities, and the 
corresponding models are constructed by the following third step.

The third step is to add $L_a$ D5 branes to each box belonging to a 
certain diagonal line of boxes, $a=1,\ldots,r$. In the singularity picture 
this corresponds to adding D branes with Chan-Paton factors given by 
$\RR_{C.P.}=\bigoplus_{i,j} L_{a(i,j)} \RR_{i,j}$, where $a(i,j)$ denotes 
the label of the diagonal passing through the box in the position $(i,j)$. 
The geometrical picture is to add $L_a$ D5 branes wrapping the $a^{th}$ 
two-cycle in the $A_{r-1}$ singularity. The field theory contains some new 
flat directions, which are mapped to the motion of these fractional branes 
along the curve of $A_{r-1}$ singularities.

The theories thus constructed satisfy automatically the condition of 
anomaly cancellation. This can be checked by noticing that at each step in 
the construction we add vector-like flavours to the gauge factors.
However, we would like to point out that the family of models we have just 
constructed is not the most general one consistent with anomaly 
cancellation. Consider for example a $3\times 3$ box model with $n$ D5 
branes in one box and zero in the others. This anomaly-free configuration 
does not belong to the class described above. 

Nevertheless, we think the family we have constructed is 
a fairly large class of models, that it includes the most general solution 
to the 
constraints in \cite{gg}, and also that some nice features of the field 
theories, to be mentioned in what follows, may allow for a study beyond 
the classical (zero string coupling) approximation.

\subsection{The one-loop beta function}

One of the simple features of this family of theories is the expression 
for the one-loop $\beta$ function of the gauge factors. Let us compute it 
first from the field theory point of view. Recall the one-loop $\beta$ 
function for a $\NN=1$ $SU(N_c)$ theory with $N_f$ (vector-like) flavours 
is proportional to $b_0=3N_c-N_f$.

In the initial configuration, all gauge groups have three flavours, and 
the one-loop $\beta$ function vanishes. After the first step, the group in 
the box $(i,j)$ has increased its rank in $N_i$ units, and its number of 
flavours increases by $N_{i-1}+N_i+N_{i+1}$, so the $b_0$ coefficient 
changes by
\beq
\Delta_1 b_0 = 2N_i - N_{i-1} -N_{i+1}.
\label{delta1}
\eeq
Observe this is the $\beta$ function of a $\NN=2$ $SU(N_i)$ theory with 
$N_{i-1}+N_{i+1}$ fundamental hypermultiplets. This theory is 
actually realized along a flat direction of the $\NN=1$ theory.

Similarly, after the second step, the group in the box at position $(i,j)$ 
has increased its number of colours in $M_j$ units, and its number of 
flavours by $M_j+M_{j-1}+M_{j+1}$. The corresponding change in the 
one-loop $\beta$ function is
\beq
\Delta_2 b_0 = 2M_j - M_{j-1} -M_{j+1}.
\label{delta2}
\eeq
Similarly, after the third step, the $\beta$ function of the group changes 
by an amount
\beq
\Delta_3 b_0 = 2L_a - L_{a-1} -L_{a+1},
\label{delta3}
\eeq
where $a$ labels the diagonal line passing through the box $(i,j)$.

The complete beta function is proportional to the sum of the three 
contributions (\ref{delta1}), (\ref{delta2}), (\ref{delta3}). The 
``sewing'' of the three $\NN=2$ theories is quite manifest in the 
structure of the beta function, and suggests it could also be understood 
in the brane pictures.

\medskip

Let us start the discussion in the brane box configurations. After the 
first step in the construction, the contribution $\Delta_1$ to the 
one-loop $\beta$ function can be understood by studying the bending of the 
NS branes, since the NS$'$ branes do not bend. As in \cite{wit4d} the 
dependence of the distance between NS branes with some energy scale (in 
our case, the vev parametrizing the Higgs branch (which is the Coulomb 
branch in the $\NN=2$ theory)) is proportional to $\Delta_1$. We stress 
that it is actually naive to assume that the dependence of the gauge 
coupling with the scale is linear, as the fact that there is only one 
direction in the NS transverse to the D5 branes seems to suggest. The 
Higgs branch is 
parametrized by the coordinate $x^5$, and also by the Wilson lines of the 
D5 brane world-volume $U(1)$'s along $x^4$. Thus, the gauge coupling 
depends on these two coordinates, and actually obeys a two-dimensional 
Laplace equation, with logarithmic solutions. This, of course, is more 
intuitive in a T-dual picture where the coordinate corresponding to the 
Wilson lines is a distance. This is achieved by T dualizing along 
$x^4$, and recovering the type IIA configurations of \cite{wit4d}.

It is now clear that the bending of the NS$'$ branes takes into account, 
in a similar way, the contribution $\Delta_2$ to the one-loop $\beta$ 
function. The complete answer, as computed from field theory, is given by 
adding these contributions. For the moment we lack a complete 
understanding of how this is accomplished in the brane picture, in 
particular because the two $\NN=2$ sub-theories have logarithmic 
dependence on different Higgs branches. We will assume this to be true, on 
the basis of simplicity, and symmetry between NS and NS$'$ branes. 

Following these lines, it is clear that the third contribution, 
$\Delta_3$, 
should be reproduced by some dynamics controlling the diagonal parameters. 
As we have mentioned, the nature of these is not clear in the brane box 
picture. Also, the adequate vevs which parametrize the relevant Higgs 
branch are not manifest. Thus, any improvement on the understanding of the 
models after step 3 requires some further knowledge about these important 
issues. 

\medskip

Let us reproduce these results in the singularity picture. After the 
second step in the construction, we have a set of fractional branes which 
can move along the curve of $A_{k-1}$ singularities. The coordinates in 
this curve parametrize the Higgs branch of the $\NN=1$ theory, or the 
Coulomb branch in the corresponding $\NN=2$ theory, and provide 
the appropriate energy scale on which the gauge couplings depend. As we 
have explained, the gauge coupling for the group arising from the $i^{th}$ 
column of boxes is encoded in the integral of the Type IIB two-forms over the 
$i^{th}$ two-cycles implicit in the $A_{k-1}$ singularity. This field 
varies over the two real dimensional Coulomb branch, and has sources 
corresponding to the fractional branes wrapped around the cycles 
intersecting the $i^{th}$ two-cycle. These sources are then the $N_{i-1}$ 
fractional branes wrapping the $(i-1)^{th}$ two-cycle, the $N_{i+1}$ 
wrapping the $(i+1)^{th}$, and also the $N_i$ wrapping the $i^{th}$. They 
are sources of charge $1$, $1$ and $-2$, respectively, as corresponds to 
the intersection numbers of the cycles. The gauge coupling thus has a 
logarithmic dependence with the parameter in the two dimensional flat 
direction, proportional to $2N_i-N_{i-1}-N_{i+1}$.

We can argue in a similar way that after the second step in the 
construction, the contribution $\Delta_2$ to the one-loop $\beta$ function 
is explained by the evolution of the gauge coupling along the Higgs branch 
parametrized by the positions of the fractional branes on the curve of 
$A_{k'-1}$ singularities. Finally, since in the singularity picture the 
diagonal parameters are manifest, one can also understand the 
contribution $\Delta_3$ that appears in the final theories, after step 3. 
It appears as the dependence of the gauge coupling with the moduli 
parametrizing the curve of $A_{r-1}$ singularities. The symmetry among the 
three types of contributions is once again manifest in the singularity 
picture, and suggest the complete contribution should be the sum of all 
three, as found in the field theory computation.

\medskip

Thus we see that this class of models allows for a nice understanding of 
the one-loop $\beta$ function in terms of several ingredients entering 
the realization using brane box constructions or branes at singularities.  
One very interesting direction of future research would be
to exploit their $\NN=2$ structure to extract exact results. It would also 
be desirable to understand the one-loop $\beta$ function in other 
anomaly-free models not belonging to this class.

\section{Final comments}

In this paper we have studied the T-duality relation between two brane 
realizations of four dimensional $\NN=1$ chiral gauge theories. In the 
absence of D branes, the map is to be understood as T-duality between 
certain grids of intersecting NS fivebranes and certain Calabi-Yau 
threefold geometries, related to $\IC^3/\Gamma$ singularities. The 
D-branes can be interpreted as probes of these configurations. We have 
shown that the simplest way to argue for this T-duality map is the study 
and comparison of the four-dimensional $\NN=1$ gauge theories that appear 
in the world-volume of these probes. Using these theories as guideline we 
have provided systematic recipes to compute the T-dual picture of a 
given one.

A satisfying result is that the T-duality relates the two known 
constructions of 
$\NN=1$ finite theories, namely the brane box models and the D3 branes at 
singularities. These theories have a number of marginal couplings. 
We have centered our interest in giving them a geometrical interpretation. 
The T-duality map has proved useful in the understanding the complete set 
of parameters. `Diagonal' parameters are not obviously realized in the 
brane box setup, but appear manifestly in the T-dual singularity picture.
Hopefully, this line of thought can lead to their appropriate 
interpretation in the brane box picture.
Another issue where the T-duality has shown its usefulness is in relating 
the different brane box configurations that give rise to the same field 
theory.

\medskip

An interesting point in our research has been the $AdS$ realization of 
the large $N$ limit of these $\NN=1$ theories. We have argued that the 
marginal operators in the field theory are correctly reproduced by 
stringy twisted sectors of the $S^5/\Gamma$ orbifold. An interesting 
feature of these fields is that they propagate on a six dimensional space 
$AdS_5\times S^1$, instead of having a ten-dimensional origin. It is an 
open question how to treat the Kaluza-Klein tower of states. A possibility 
is studying Type IIB supergravity on smooth ALE spaces (times a circle) in 
presence of the RR four-form background. A more practical point of view, 
along the lines of \cite{fz}, would be to use the appropriate 
$\NN=4,2$ five-dimensional gauged supergravity. 

\medskip

Finally, we have shown how the T-duality extends to theories which are not 
conformal. These theories are easily realized in the brane box picture, 
placing different numbers of D5 branes on each box. We have argued that 
these configurations map to fractional branes generically stuck at the 
singularity. We have also shown how to determine the Chan-Paton matrices 
for these D3 brane, for a given a brane box configuration. An interesting 
point is that anomaly cancellation in the field theory imposes some 
restrictions on the possible Chan-Paton matrices. Presumably, the anomaly 
cancellation follows from some consistency condition on the construction 
of the orbifold. 

We have also presented a quite large family of anomaly-free models, 
obtained by ``sewing'' together several $\NN=2$ models. A subset of this 
theories provides the most general solution to the ``sum of diagonals'' 
rule, but the complete family is more general, violating that condition in 
many cases. However the construction in the singularity picture is very 
symmetric and suggests the consistency of these configurations even at the 
quantum level. 

Even though these theories are $\NN=1$ supersymmetric, there are Higgs 
branches along 
which $\NN=2$ is restored. The theories have a very simple one-loop 
$\beta$ function, which we have (partially) explained in terms of the 
brane pictures. 

\centerline{Acknowledgements}

We would like to thank Michael Douglas, Luis E.~Ib\'a\~nez, Anton 
Kapustin, Jaemo Park and Matthew Strassler for fruitful discussions. A.~U. 
is grateful to M.~Gonz\'alez for advice and support.
A. H. is supported in part by National Science Foundation grant
NSF PHY-9513835. This research was supported in part by
NSF Grant PHY-9407197 through the Institute for Theoretical Physics in
Santa Barbara. A. H. would like to thank the hospitality of the ITP while
this work was being completed.  The work of A.U.~was
supported by the Ram\'on Areces Foundation, and partially by the CICYT 
under grant AEN97-1678. 

\bibliography{singu}
\bibliographystyle{utphys}
\end{document}